\documentstyle [12pt,psfig] {article}

\newcommand{\nwc}{\newcommand}
\nwc{\hyp} {\hyphenation}
\nwc{\be}  {\begin{equation}}
\nwc{\ee}  {\end{equation}}
\nwc{\ba}  {\begin{array}}
\nwc{\ea}  {\end{array}}
\nwc{\bdm} {\begin{displaymath}}
\nwc{\edm} {\end{displaymath}}
\nwc{\bea} {\be\ba{lcl}}
\nwc{\eea} {\ea\ee}
\nwc{\bda} {\bdm\ba{lcl}}
\nwc{\eda} {\ea\edm}
\nwc{\bc}  {\begin{center}}
\nwc{\ec}  {\end{center}}
\nwc{\ds}  {\displaystyle}
\nwc{\bmat}{\left(\ba}
\nwc{\emat}{\ea\right)}
\nwc{\non} {\nonumber}
\nwc{\hph} {\hphantom}
\nwc{\qq}  {\qquad}
\nwc{\lra} {\longrightarrow}
\nwc{\ra}  {\rightarrow}
\nwc{\Ra}  {\Rightarrow}

\nwc{\rng} {\rangle}
\nwc{\lng} {\langle}

\nwc{\lmt} {\longmapsto}
\nwc{\sq}  {\sqrt}
\nwc{\prl} {\partial}
\nwc{\prlb}{\bar{\prl}}
\nwc{\fc}  {\frac}
\nwc{\kr}  {\kern}
\nwc{\iy}  {\infty}
\nwc{\ol}  {\overline}
\nwc{\hm}  {\hspace{3mm}}
\nwc{\Er}   {{\rm E}}
\nwc{\tr} {\rm Tr}
\nwc{\lf} {\left}
\nwc{\ri} {\right}
\nwc{\lm} {\limits}
\nwc{\lb} {\lbrack}
\nwc{\rb} {\rbrack}
\nwc{\ov} {\over}
\nwc{\ovx}{\over\textstyle}
\nwc{\til}{\tilde}
\nwc{\pr} {\prime}
\nwc{\st} {\strut}
\nwc{\vs}  {\vskip}
\nwc{\noa} {\noalign}
\nwc{\scr}  {\scriptstyle}
\nwc{\tx}  {\textstyle}
\nwc{\scs} {\scriptscriptstyle}
\nwc{\Th} {\Theta}
\nwc{\th} {\theta}
\nwc{\vth} {\vartheta}
\nwc{\eps}{\epsilon}
\nwc{\si} {\sigma}
\nwc{\Gm} {\Gamma}
\nwc{\gm} {\gamma}
\nwc{\bt} {\beta}
\nwc{\La} {\Lambda}
\nwc{\la} {\lambda}
\nwc{\om} {\omega}
\nwc{\Om} {\Omega}
\nwc{\dt} {\delta}
\nwc{\Si} {\Sigma}
\nwc{\Dt} {\Delta}
\nwc{\al} {\alpha}
\nwc{\vph}{\varphi}
\nwc{\Sis} {\Si^{(s)}}
\nwc{\sis} {\si^{(s)}}
\nwc{\Sims}{\Sigma^{(-s)}}
\nwc{\Us}  {U^{(s)}}
\nwc{\sipl} {\sigma^+}
\nwc{\simi} {\sigma^-}
\nwc{\sd}  {\dot s}
\nwc{\Id}  {{\bf 1}}
\nwc{\Sc}  {{\cal S}}
\nwc{\Cr}  {{\cal R}}
\nwc{\Cd}  {{\cal D}}
\nwc{\Oc}  {{\cal O}}
\nwc{\Cc}  {{\cal C}}
\nwc{\Of}  {{\cal O}_f}
\nwc{\Oft} {{\cal O}_{f_2}}
\nwc{\Ofo} {{\cal O}_{f_1}}
\nwc{\Pc}  {{\cal P}}
\nwc{\Mc}  {{\cal M}}
\nwc{\Ec}  {{\cal E}}
\nwc{\Fc}  {{\cal F}}
\nwc{\Hc}  {{\cal H}}
\nwc{\cK}  {{\cal K}}
\nwc{\Fcp} {{\cal F}^\pr}
\nwc{\Hcp} {{\cal H}^\pr}
\nwc{\Xc}  {{\cal X}}
\nwc{\Gc}  {{\cal G}}
\nwc{\Zc}  {{\cal Z}}
\nwc{\Nc}  {{\cal N}}
\nwc{\fca} {{\cal f}}
\nwc{\xc}  {{\cal x}}
\nwc{\Ac}  {{\cal A}}
\nwc{\Bc}  {{\cal B}}
\nwc{\Uc} {{\cal U}}
\nwc{\Vc} {{\cal V}}
\nwc{\rh}  {{\hat r}}

\nwc{\nnn} {\nonumber \vspace{.2cm} \\ }
\nwc{\hb}  {\bar h}
\nwc{\xb}  {\bar x}
\nwc{\yb}  {\bar y}
\nwc{\zb}  {\bar z}
\nwc{\wb}  {\bar w}
\nwc{\Ob}  {\bar O}
\nwc{\Yb}  {\bar Y}
\nwc{\cv} {{\vec c}}
\nwc{\ph} {{\hat p}}
\nwc{\wh} {{\hat w}}
\nwc{\qh} {{\hat q}}
\nwc{\xh} {{\hat x}}
\nwc{\php} {{\hat p}^\pr}
\nwc{\whp} {{\hat w}^\pr}
\nwc{\qhp} {{\hat q}^\pr}
\nwc{\xhp} {{\hat x}^\pr}
\nwc{\bp}  {b^\pr}
\nwc{\ft} {{\til f}}
\nwc{\ct} {{\til c}}
\nwc{\tit}{{\til t}}
\nwc{\wt} {{\til w}}
\nwc{\Pb} {{\bf  P}}
\nwc{\Zb} {{\bf  0}}
\nwc{\Eb} {{\bf  1}}
\nwc{\Qb} {{\bf  Q}}
\nwc{\Bb} {{\bf  B}}
\nwc{\Tb} {{\bf  T}}
\nwc{\qqd}  {\qq .}
\nwc{\diag} {{\rm diag}}
\nwc{\inv}  {{\rm inv}}
\nwc{\mod}  {{\rm mod}}
\nwc{\clr}  {{\rm cl}}
\nwc{\qur}  {{\rm qu}}
\nwc{\Pro}   {{\rm P}}
\nwc{\Yr}   {{\rm Y}}
\nwc{\sof}  {{\st 1 \ov \sqrt{|\Of|}} }
\nwc{\sofo} {{\tx 1 \ovx \sqrt{\tx |\Ofo|}} }
\nwc{\soft} {{\tx 1 \ovx \sqrt{\tx |\Oft|}} }
\nwc{\hal} {\frac{1}{2}}
\nwc{\tpi}  {2\pi i}
\nwc{\gpr}  {g^\pr}
\nwc{\zot}  {z_{12}}
\nwc{\zbot} {\zb_{12}}
\def\KK{{\rm I\kern -.2em  K}}
\def\NN{{\rm I\kern -.16em N}}
\def\RR{{\rm I\kern -.2em  R}}
\def\ZZ{Z \kern -.43em Z}

\def\QQ{{\rm \kern .25em
             \vrule height1.4ex depth-.12ex width.06em\kern-.31em Q}}
\def\CC{{\rm \kern .25em
             \vrule height1.4ex depth-.12ex width.06em\kern-.31em C}}

\newcommand{\sth}{string theory}

\newcommand{\fth}{field theory}

\def\ZZZ{Z\kern -0.31em Z}

\newcommand{\sect}[1]{ \section{#1} \setcounter{equation}{0} }

\newcommand{\p}{\partial}
\newcommand{\f}{\hat{f}}
\newcommand{\w}{\hat{v}}
\newcommand{\m}{\hat{p}}

\newcommand{\eqr}{ \begin{eqnarray}}
\newcommand{\rqe}{ \end{eqnarray}}
\newcommand{\eq}{\begin{equation}}
\newcommand{\qe}{\end{equation}}
\newcommand{\req}[1]{(\ref{#1})}

\newcommand{\cu}{{\cal U}}
\newcommand{\Ga}{\Gamma}

\newcommand{\g}{\mbox{$g_k(\hat{P}_{L},\hat{P}_{R})$}}

\newsavebox{\nnin} \sbox{\nnin}{$\hspace{1mm}\in\kern -.8em /
                   \hspace{1mm}$}
\newcommand{\nin}{\usebox{\nnin}}

\newcommand{\sub}{\subset}
\newsavebox{\nnsub} \sbox{\nnsub}{$\hspace{1mm}\sub\kern -.9em /
            \hspace{1mm}$}
\newcommand{\nsub}{\usebox{\nnsub}}

\begin{document}

\begin{titlepage}

\title{Higher Twisted Sector Couplings of $Z_N$
       Orbifolds\thanks{Supported by the Deutsche Forschungsgemeinschaft}}

\author{{\sc J. Erler$^{1,2}$}\and {\sc D. Jungnickel$^{1,2}$} \and
        {\sc M. Spali\'nski$^2$}\thanks{
       Alexander von
       Humboldt Fellow. On leave from the Institute of Theoretical
       Physics, Warsaw University}
       \and {and \sc S. Stieberger$^2$\ \ \ \ } \\ \\ \\
       {\em $^1$Max--Planck--Institut
        f\"ur Physik} \\
       {\em ---Werner--Heisenberg--Institut---}\\
       {\em P.O. Box 401212, D--8000 M\"{u}nchen, Germany}\\ \\
       {\em $^2$Physik Department} \\
       {\em Technische Universit\"at M\"unchen} \\
       {\em D--8046 Garching, Germany}}
\date{}
\maketitle

\begin{picture}(5,2.5)(-350,-440)
\put(12,-90){MPI--Ph/92-56}
\put(12,-105){TUM--TH--142/92}
\put(12,-123){June 1992}
\end{picture}

\thispagestyle{empty}

\ \\

\begin{abstract}
  We derive the basic correlation functions of twist fields coming from
  arbitrary twisted sectors in symmetric $Z_N$
   orbifold conformal field theories, keeping all the
  admissible  marginal perturbations, in particular
  those corresponding to
  the antisymmetric tensor background field. This allows a thorough
  investigation of modular symmetries in this type of
  string compactification. Such a study is
  explicitly carried out for the group generated by
  duality transformations. Thus, apart from being
  of phenomenological use, our couplings are also interesting from the
  mathematical point of view as they represent automorphic
  functions for a large class of discrete groups.
\end{abstract}

\end{titlepage}

\sect{Introduction}

It was recognized  some years ago that \sth\ could be a viable
framework for gravity as well as other interactions. At that time it was
already clear that compared to ordinary local \fth\
it is much more restrictive. This lead
to the belief that this framework would be restrictive enough to select a
more or less unique ``Theory of Everything'' \cite{chsw}.
It has since become clear
that the existing (perturbative) formulation does, however,
admit a great deal of
arbitrariness concerning the vacuum state of string theory, which is
of course reflected in the low energy effective field theory.
One of the
tasks of the string phenomenologist is to determine which field theories
can be interpreted as a low energy description of a string
compactification.

The arbitrariness inherent in the presently available formulation of
\sth\ is very poorly understood. In a generic
compactification it manifests itself in
the dependence of physical quantities on a set of parameters
--- the moduli~\cite{dvv}.  It is not known if (and how)
\sth\ can determine the values of these parameters.
However, a first attempt in this direction
is based on the assumption that discrete symmetries
in moduli space (see below) survive after
taking into account non--perturbative
string effects~\cite{ft}.
This assumption limits the form of any
possible non--perturbative contribution to the
superpotential, which can then
lead to the lifting of the vaccuum degeneracy.

A tractable example of a nontrivial
family of string vacua is the class of orbifold
compactifications \cite{soo}, where the moduli are accessed easily as
the radii and angles specifying the underlying torus.
In this class of
models it is possible to calculate practically all quantities of physical
interest and study their moduli dependence. Furthermore these models
possess exiting phenomenological properties~\cite{INQ}
and three generation
models with gauge group $SU(3) \times SU(2) \times U(1)^n$ are easily
constructed~\cite{IKNQ}.

The work presented here concentrates on the calculation of
correlation functions in the orbifold conformal field theory,
with the aim of reproducing their complete dependence on the
moduli. In the case of the three twist correlation functions
(directly related to the Yukawa couplings in the effective
\fth), the moduli dependence is of exponential type, due to the
presence of world sheet instantons \cite{dsww,dix}. This could be
used to discuss, for example, quark mass hierarchies
\cite{cm,cgm}.
Indeed, any phenomenological studies based on orbifold
compactifications will require the knowledge of these correlation
functions.

Correlation functions of
twist fields from the lower twisted sectors
were first computed by Dixon,
Friedan, Martinec, and Shenker \cite{dix}, and (independently) by Hamidi
and Vafa \cite{hamidi}. In these papers the basic techniques for
orbifold calculations were developed. In both cases the twisted Narain
compactification
\cite{gso,N,nsw} consisted of a purely metric background.
Antisymmetric tensor backgrounds in the case of a two--dimensional target
space were treated in \cite{rp}, and subsequently
the annihilation couplings \cite{ejlm}
and particular Yukawa couplings \cite{ejl} were derived in a general
background in higher dimensions.

The extension to an arbitrary choice of higher sector twist fields
was initiated in \cite{bur} for the case of a purely metric
background. Examples of couplings
for six--dimensional $Z_N$ heterotic Coxeter
orbifolds \cite{gso} have recently been worked out in \cite{casas}, again
with just a metric background and without proper normalization.
In the present article all three point correlation functions for bosonic
$Z_N$ orbifolds are determined as a function of all background
deformations present. We calculate all couplings admitted by the orbifold
selection rules, i.e.\ not only those relevant for the heterotic string.

Some of our results related to higher sector twist field
configurations do not agree with the findings of \cite{bur}.
These discrepancies show up in cases
when none of the twist sector numbers
is a multiple of the others.
The source of this lies in the complete set of
global monodromy conditions which must be taken
into account in determining
the admissible
world sheet instantons. What we find is that in
some cases not all the world
sheet instantons included in previous calculations actually contribute.

Having
the complete moduli dependence of the couplings under control
makes it possible
to investigate discrete symmetries in moduli space. These
symmetries, below referred to as modular symmetries, relate
compactifications which are characterized
by different geometries (as specified by the values of the
moduli). The best known symmetry of this kind is target space
duality \cite{ky,dvv,giveon}. In the context of orbifolds, modular
symmetries are easily identified in the spectrum of the
theory \cite{dhs}--\cite{ejn}.
It is clearly important to know whether they are also
respected by the interactions of compactified
heterotic string models. Many examples are known where this
is the case \cite{dvv}, \cite{rp}--\cite{ejl},
\cite{nilles}--\cite{cogp}. It would be
rather surprising if there were interactions which break duality
invariance. One may even be tempted to consider duality
invariance as a consistency check on string calculations. This point of
view was articulated in \cite{sjls},
where some of the results derived below
have been presented.

The organization of this paper is as follows:
in section~\ref{fourpointcorrelation} we
derive the four twist field correlator for an arbitrary twist field
configuration
with the full background dependence.
It is shown there that in general one
has to
consider three different contours in order to obtain the complete
global monodromy information.
This is to be compared with the cases treated
in \cite{dix,rp,ejl}, where it is sufficient to take just two contour loops.

In section~\ref{YY} we perform the factorization of this
four point function into the s--channel, i.e.\ the channel with untwisted
intermediate states.
In doing so we recover the twist--anti--twist annihilation
coupling, which can be compared with the one derived in~\cite{ejlm}.
Factorization into the u--channel is the subject of
section~\ref{YYY}, where all Yukawa couplings are calculated.
The correctness of
the summation range over world sheet instanton contributions is confirmed
by a direct calculation of these couplings sketched in
appendix~\ref{direct}.

The peculiar situation where the orbifold
twist leaves some fixed directions
is analyzed in section~\ref{fixtori}.
A general prescription is given for
handling such cases.

Sections~\ref{sec6} and~\ref{sec7} are devoted to the study of duality
invariance of the theory.
Having found the complete set of couplings we were able to show that the
theory is invariant under a large group generated by duality
transformations. As found earlier in simpler cases
\cite{nilles}, \cite{rp}--\cite{ejl} the background
transformations have to be accompanied by transformations of the twisted
sector ground states.
In the general case considered here this twist field
transformation depends on both the element of the duality group acting on
the background and on the background itself. We take the fact that the
theory turns out to be duality
invariant as a consistency check on the calculation of the couplings
compared to the results of \cite{bur}.

\sect{Four twist correlations with general B--field}
\label{fourpointcorrelation}

The starting point for the construction of
two--dimensional orbifold conformal
field theories is the Euclidean linear sigma model action
for $d$ bosonic coordinate
fields \cite{nsw}
\be
   S_E [G,B,X] = \frac{1}{\pi \alpha'} \int \ dz \ d \bar z \ \bar \p X^t
   (z,\ol{z}) \ (G+B)\ \p X(z,\ol{z})\; ,
   \label{action1}
\ee
where $B$ denotes the constant\footnote{The term ``constant''
means, that $B$ neither depends on the fields $X^i(z,\ol{z})$ nor
on the world sheet coordinates $z$, $\ol{z}$.}
antisymmetric background tensor field and
$\alpha'$ is the inverse string tension, which will be fixed in
the following according to $\alpha'=2$. The
$d$--dimensional compact part of the string target manifold\footnote{
We assume $d$ to be an even positive
integer. A method for generalizing our results to odd $d$ may be found
in \cite{ejl}.}, described by the
{\em coordinate fields} $X^i(z,\bar z)$,
is assumed to be a toroidal orbifold $O_d :=\RR^d /S$,
where $S$ denotes the so called {\em space group}
$$ S=\{(\Theta ,w)| w \in \Lambda \} \; ,$$
which is a discrete subgroup of the Euclidean group of $\RR^d$.
$\Lambda$ denotes a $d$--di\-men\-sional lattice and the allowed $\Theta$ are
discrete rotations of $\RR^d$, which build the {\em point group} $P$
of the orbifold.
Except possible fixed points under $P$ the orbifold
target space
is flat and we may choose a coordinate system such that its
(constant) metric  $G=\frac{1}{2}{\bf 1}$.

More precisely,
the construction $O_d :=\RR^d /S$ means, that each point $x\in\RR^d$ has
to be identified with $\Theta x +2\pi w$ if $(\Theta ,w)\in S$. This
implies closed string boundary conditions
\be
 X(e^{2\pi i}z,e^{-2\pi i}\bar z)=(\Theta X)(z,\bar z) + 2\pi w
 \label{bouncond}
\ee
for the coordinate fields.
In this paper we will restrict ourselves to symmetric
$Z_N$ orbifolds \cite{soo,gso,ek}, for which the point group is generated
by a single twist $\Theta$ of order $N$ which
acts symmetrically on the right and left moving parts\footnote{
The equation of motion
$\p \bar \p X(z,\bar z)=0$ following from (\ref{action1}) is
solved by the general ansatz $X(z,\bar z)=X_R(z)+X_L(\bar z)$,
where $X_R(z)$ and $X_L(\bar z)$ are referred
to as the right and the left
moving parts of $X(z,\ol{z})$, respectively.}
of $X(z,\bar z)$.

We block--diagonalize the twist matrix $\Theta$ by an element of
$O(d)$ without altering our choice of $G$
\bdm \Th = \lf(\ba{ccc}
      \vartheta (\al_1) &        &         \\
                & \ddots &         \\
                &        & \vartheta (\al_{d\ov 2} )
                           \ea \ri) ,\;\;
     \vartheta (\al_j)=\lf(\ba{rr}  \cos (2\pi \al_j) & \sin (2\pi \al_j) \\
     -\sin (2\pi \al_j) &  \cos (2\pi \al_j) \ea \ri)
\edm
with $N\al_j\in\{1,\ldots,N-1\}$.
The a priori possible cases $N\al_j$ $=$
$0$ mod $N$, which correspond to fixed tori of the orbifold
will be discussed in section \ref{fixtori}.

It then proves to be useful to switch from the $d$ real coordinate fields
$X^j(z,\bar z)$ to $d/2$ complex variables
\bea
 Y^j(z,\bar z) &:=& X^{2j-1}(z,\bar z) + i X^{2j}(z,\bar z) \nnn
 \bar Y^j(z,\bar z) &:=& X^{2j-1}(z,\bar z)
 - i X^{2j}(z,\bar z) \ \ ,\  j=1,
 \cdots, d/2 \; . \label{complex}
\eea

Equipped with these definitions we now proceed
with the calculation of the correlation function
\be
 Z_{\{f_{i}\}}(x,\bar x)=\lim_{|z_{\infty}| \rightarrow
 \infty}|z_{\infty}|^{4h_l^\si}\ \lng \sigma_{f_1}^{-k}(0,0)
 \sigma_{f_2}^{k}(x,\bar x)\sigma_{f_3}^{-l}(1,1)
 \sigma_{f_4}^{l}(z_{\infty},\bar z_{\infty}) \rng \; .
 \label{correlation}
\ee

It will be used to derive operator product coefficients of the
underlying conformal field theory (CFT) via factorization.
Here $\sigma^k_f$ denotes a twist field of the $k$--th
twisted sector corresponding to the fixed point $f$
satisfying\footnote{Note, that a factor $2 \pi$ is included in the
definition of $f$.} $\Th^k f=f$ ${\rm mod}$ $\La$,
which is defined via its operator product expansion (OPE) with the
coordinate differentials:
\be \begin{array}{rclrl}
 \ds{\p Y^j(z,\bar z)\  \sigma^{k}_{f}(w,\bar w)}
 & = & \ds{(z-w)^{-(1-k_j)}
 }&\ds{ (\tau_j)^{k}_f(w,\bar w)}+\ldots & \nnn
 \ds{\p \bar Y^j(z,\bar z)\
 \sigma_{f}^{k}(w,\bar w)} & = & \ds{(z-w)^{-k_j}
 }&\ds{ (\tau'_j)^{k}_f(w,\bar w)}+\ldots & \nnn
 \ds{\bar \p
 Y^j(z,\bar z)\  \sigma_{f}^{k}(w,\bar w)} & = & \ds{(\bar z-\bar
 w)^{-k_j}}&\ds{ (\tilde{\tau}_j^\prime )^{k}_f(w,\bar w)}+\ldots & \nnn
 \ds{\bar \p \bar Y^j(z,\bar z)\  \sigma_{f}^{k}(w,\bar w)}
 & = & \ds{(\bar z-\bar w)^{-(1-k_j)}}
 &\ds{(\tilde{\tau}_j)^{k}_f(w,\bar w)}+\ldots \; , &
 \end{array} \label{twistOPE}
\ee
with  $k_j :=[k\alpha_j]\in (0,1)$ . These relations also
serve as definitions of four different types of
so called {\em excited twist fields\/} on the right hand side of
(\ref{twistOPE}).

We assume that $k_j,l_j\neq 0$ mod $N$, i.e. neither $\Th^k$ nor $\Th^l$
should possess fixed directions. This requirement is generally stronger
than the above assumption $N\alpha_j\neq 0$ mod $N$, which is a
consequence of it. Of course, (\ref{correlation}) is easily
evaluated for the case that $\Th^k$ or $\Th^l$
leave some of the coordinates
invariant. The whole correlator may be written as a product over all
two--dimensional
complex subplanes\footnote{An a priori possible coupling of twisted and
untwisted coordinate fields via the $B$--field is easily seen to
be forbidden by the requirement $[B,\Th ]=0$ to be discussed later.}
corresponding to (\ref{complex}) as is obvious from its path
integral representation. The contributions from invariant
two--dimensional subplanes
are actually two--point functions of twist fields and (take for
instance $k_j =0$) normalized according to
\be \lng {\bf 1}\; \si^{-l_j}_{f_3^j}(1,1)
    \si^{l_j}_{f_4^j}(z_\infty,\zb_\infty)\rng =
    |z_\infty |^{-4h_{l_j}^\si}\dt_{0\in f_4^j - f_3^j + \La^j}\; ,
    \label{norm}
\ee
where $\La^j$ denotes the projection of $\La$ on the $j$--th
two--dimensional subplane and $h_{l_j}^\si$ is given in (\ref{hkj}).
Nevertheless
the possibility of fixed tori is interesting because they allow for
nontrivial winding and momenta w.r.t.\ the invariant directions.
This results in a drastic modification of the spectrum of such models,
which together with the associated additional couplings will be discussed
in section \ref{fixtori}.

The adjoint of $\si^k_f$
or {\em anti--twist field} is $\sigma_f^{-k}$
and their conformal weights are given by
\be
 h_k^\si=\bar h_k^\si =\sum_{j=1}^{d/2}h_{k_j}^\si  =
 \sum_{j=1}^{d/2} \frac{1}{2} k_j(1-k_j)\; .
 \label{hkj}
\ee
Actually for higher twisted sectors ($k>1$) these twist fields are not
twist invariant and may therefore not be considered as physical fields.
It is however possible to construct twist invariant linear combinations
of the above twist fields, as explained in \cite{rp}. They are given by
\be
   \Sigma_{\Oc_{k;f}}^{k}=\frac{1}{|\Oc_{k;f}|}\
   \sum_{f' \in \Oc_{k;f}} \sigma^{k}_{f'}\; ,
\label{twistinv}
\ee
where $\Oc_{k;f}$ denotes the {\em orbit} of the fixed point $f$
under $\Th^k$:
\bdm \Oc_{k;f}\equiv \left\{ \Th^j f\;\; {\rm mod}\;\;
     \La\ |\ 0\leq j<k \right\}\; .
\edm
Since the physical twist fields $\Sigma_{\Oc_{k;f}}^{k}$
are linear combinations
of the non--twist invariant $\si_f^k$, we may for
simplicity only consider the
latter ones in the following. All our results are easily extended to the
$\Sigma_{\Oc_{k;f}}^{k}$.

To each twist field $\sigma_f^k$ one may associate a
conjugacy class $[\Theta^k,$ $(1-\Theta^k)(f+\La)]$ of space group
elements. The {\em space group
selection rule} for orbifold
correlators \cite{dix,hamidi} now states that any correlation function
of twist fields vanishes, if the product of their corresponding
conjugacy classes does not contain the space group identity element
$({\bf 1},0)$. In particular this means that the total ``net twist''
of any correlator has to vanish.
This is the so called {\em point
group selection rule}, which was taken into account in
(\ref{correlation}).
The space group selection rule for this correlation function
reduces to\footnote{Note that there are
further selection rules in the case of supersymmetric or heterotic
string theories \cite{font,kob}.}
($f_{ij}:=f_i -f_j$)
\be
 (1-\Th^k)(f_{21}-T_1)+(1-\Th^l)(f_{43}-T_2)=0
 \label{sgsr1}
\ee
for some $T_1$, $T_2$ $\in$ $\La$.

To evaluate (\ref{correlation}) we use the techniques
developed in \cite{dix}.
We split the
coordinate fields $X^j$ into a classical and a quantum piece:
\bdm X^j(z,\bar z)=X_{cl}^j(z,\bar z)+X_{qu}^j(z,\bar z)\ ,
\edm
where $X_{cl}^j$ denotes a solution of the classical equation of motion
$\p \bar \p X^j(z,\bar z)=0$ following from
(\ref{action1}) and $X_{qu}^j$
describes the quantum fluctuations around these
instantons. In the presence
of a twist field, for simplicity located at the origin, the classical
field must obey the full boundary condition (\ref{bouncond})
whereas the quantum part only feels the {\em local} monodromy:
$$ X_{qu}(e^{2\pi i}z,e^{-2\pi i}\bar z)=(\Theta X_{qu})(z,\bar z)\; .$$

Due to the fact that (\ref{action1}) is bilinear
in the fields $X^j$,
the path integral representation of (\ref{correlation})
naturally splits into two factors, one representing
the solutions to the classical equation of motion
for $X$, the other describing
the {\em quantum} fluctuations around these instantons. This means
\be
 Z_{\{f_{i}\}}(x,\bar x) =Z_{qu}(x,\bar x)\
 \sum_{X_{cl}}
 e^{-S_E [G,B,X_{cl}]} \; ,
\ee
where the instanton solutions $X_{cl}^j$ are subject to the full boundary
conditions imposed by the four twist fields in (\ref{correlation}).
In view of the local monodromy conditions implied by
(\ref{twistOPE}) we are left with the most general Ansatz \cite{bur}
\be
 \begin{array}{lcllcl}
 \p Y^j_{cl}(z) &=& b^j \omega_{k_j,l_j}(z) \;\;\; & \bar \p
 \bar{Y}^j_{cl}(\bar{z})
 &=& \bar{b}^j \bar{\omega}_{k_j,l_j}(\bar{z}) \nnn
 \bar \p Y^j_{cl}(\bar{z})
 &=& c^j \bar{\omega}_{1-k_j,1-l_j}(\bar{z})\;\;\;
 & \p \bar
 Y^j_{cl}(z) &=& \bar c^j \omega_{1-k_j,1-l_j}(z) \; ,
 \end{array}
 \label{Ansatz}
\ee
where we defined
\be \begin{array}{lcr}
 \ds{\omega_{k_j,l_j}(z)} &:=&
 \ds{z^{-k_j}(z-x)^{k_j-1}(z-1)^{-l_j}(z-z_{\infty})^{l_j-1}} \nnn
 \ds{\omega_{1-k_j,1-l_j}(z)} &:=& \ds{z^{k_j-1}(z-x)^{-k_j}
 (z-1)^{l_j-1}(z-z_{\infty})^{-l_j}}\; . \end{array}
 \label{cutdifferentials}
\ee

So far we only made use of the {\em local}
monodromy conditions imposed by
the presence of the four twist fields in (\ref{correlation}).
In order to employ their {\em global} properties
we choose closed contours, each containing
two of the twist fields,
in such a way that the enclosed net twist is zero.
When transported around such loops the
instantons are shifted by some coset vector, which is determined
by multiplying the space group conjugacy classes
$[\Th^k,(1-\Th^k)(f+\La)]$ corresponding to
the encircled twist fields.
It was also pointed out in \cite{dix} that for $k=l=1$
all global information may
be obtained by considering only {\em two} net twist zero contours:
$C_1$ enclosing the points $0$, $x$ and $C_2$ encircling
$x$, $1$. These considerations easily generalize to arbitrary $k=l$.

For $k\neq l$ the situation changes in two respects: First, as
described in \cite{bur}, the loop $C_2$ in order to be closed
now must surround the points $x$ $p$--times
and $1$ $q$--times, where
$pk=ql$ mod $N$  and $p,q$ are taken to be
the smallest positive integers with this
property. Secondly, there exists a {\em third independent\/} loop
$C_3$ enclosing the world sheet punctures $1$ and $\infty$.
These three loops are drawn in figure \ref{C1C2C3} on the cut
complex plane for the case $p=2$, $q=1$.
Actually the branch cuts are a property of
the instanton solutions (\ref{Ansatz}), whereas the world sheet
is the cut--free compactified complex plane $\CC\cup\{\infty\}$.
However, sometimes it proves to be useful to assume that the instantons
are single valued functions on an
appropriately cut world sheet \cite{hamidi}.
This point of view is chosen in figure \ref{C1C2C3}.

The third net twist zero loop $C_3$ is the source of the discrepancy
between our results and those of \cite{bur} for $k\neq l$.
We will see in the following, that taking $C_3$ into account is
(among other things) necessary to obtain a crossing symmetric result
for the four--point function (\ref{correlation})\footnote{One might
suspect that there is a fourth independent net twist zero loop
$C_4$, encircling $0$ $p$--times and $\infty$ $q$--times.
But, when solving the global monodromy conditions for this loop,
it turns out, that no further information is gained by $C_4$
or any other loop.}.

We therefore have to evaluate three ($a=1,2,3$) sets of generically
independent {\em global} monodromy conditions:

\begin{figure}
\psfig{figure=stem.ps,height=95mm,width=115mm}
\caption{The loops $C_1, C_2$ and $C_3$ for the case $p=2, q=1$.}
\label{C1C2C3}
\end{figure}

\be \begin{array}{rcl}
 \ds{\triangle_{C_a} Y^j_{cl}}&\equiv&\ds{ \oint_{C_a} dz \p Y_{cl}^j
 +\oint_{C_a} d\bar z \bar \p
 Y_{cl}^j = 2\pi v_a^{j}} \nnn
 \ds{\triangle_{C_a} \bar Y^j_{cl}} &\equiv&\ds{ \oint_{C_a} dz
 \p \bar Y^j_{cl} +
 \oint_{C_a} d \bar z \bar \p \bar Y^j_{cl} =2\pi \bar v_a^{j}} \; .
 \end{array} \label{clmond}
\ee
The complex vectors $v^j$ are defined via
$v^j := v^{(2j-1)} + i \ v^{(2j)}$, where $v^{(j)}$ denotes
the $j$--th component of the $d$--dimensional real vector $v$.
They are restricted by
\bea
 v_1 &\in& F_1 := (1-\Th^k)(f_{21}+\La ) \\[3mm]
 v_2 &\in& F_2 := (1-\Th^{pk})(f_{23}+\La ) \\[3mm]
 v_3 &\in& F_3 := (1-\Th^l)(f_{43}+\La ) \; .
 \label{mv}
\eea

For given $v_1$, $v_2$, $v_3$ (\ref{clmond}) is to be understood as
a system of $3\cdot\fc{d}{2}$ complex linear equations for the
$2\cdot\fc{d}{2}$ complex instanton coefficients $b^i$, $c^i$.
Each solution, parametrized by $v_1$, $v_2$, $v_3$,
yields one instanton. If, as mentioned before, for $k\neq l$ all
three loops give independent global monodromy conditions, this system
is obviously over--determined. This is a new feature of the
case $k\neq l$ compared to $k=l$, which is due to the third
independent loop. For $k=l$ only $C_1$ and $C_2$ give independent
monodromy relations and one obtains $2\cdot\fc{d}{2}$ complex
equations for the same number of unknown complex quantities \cite{dix}.

The solution for $a=1,2$ may be inferred from \cite{bur}. It is
given by
\be b^{j} = b_1^j v_1^{j}\ +\ b_2^j v_2^{j}\; ,
    \;\;\; c^{j}=c_1^j v_1^{j}\ +\ c_2^j v_2^{j}
    \label{Instcoef}
\ee
with
\be
\begin{array}{l}
 \ds{b_1^j = -i(1-\bar x)^{l_j-k_j} e^{i\pi (2k_j-l_j)}
 \fc{J_{2j} \bar H_{2j}(1-\bar x)}{I_j(x,\bar x)}} \nnn
 \ds{b_2^j =-i e^{i\pi [(2-p)k_j - l_j]} \fc{\bar G_{1j}(\bar x)}
 {I_j(x,\bar x)}} \fc{\pi}{\sin(pk_j\pi )} \nnn
 \ds{c_1^j =-i(1-x)^{k_j-l_j} e^{i\pi (2k_j-l_j)}
 \fc{J_{1j} H_{1j}(1-x)}{I_j(x,\bar x)}} \nnn
 \ds{c_2^j =i e^{i\pi [(2-p)k_j-l_j]}
 \fc{G_{2j}(x)}{I_j(x,\bar x)}} \fc{\pi}{\sin(pk_j\pi )}\; .
 \end{array} \label{bbcc}
\ee
Here we used the following abbreviations for hypergeometric and gamma
functions:
\bdm
 \ds{G_{1j}(x):=F(k_j,1- l_j;1;x)}\;\; ,\;\; G_{2j}(x):= F(1-k_j,l_j;1;x)
\edm
\bda
 H_{1j}(x) &:=& \ds{F(k_j,1-l_j;1+k_j-l_j;x)} \nnn
 H_{2j}(x) &:=& \ds{F(1-k_j,l_j;1+l_j-k_j;x)}
\eda
\bdm
 J_{1j}:=\frac{\Gamma(k_j) \Gamma(1-l_j)}{\Gamma(1+k_j-l_j)}\;\; ,\;\;
 J_{2j}:=\frac{\Gamma(1-k_j)\Gamma(l_j)}{\Gamma(1+l_j-k_j)}
\edm
\bdm I_j(x,\xb ):=
     J_{2j}\ol{G}_{1j}(\xb )H_{2j}(1-x)+
     J_{1j}G_{2j}(x)\ol{H}_{1j}(1-\xb ) \; ,
\edm
which satisfy the useful identities
\bdm
    J_{1j}\ G_{2j}(x)\  H_{1j}(1-x)
    =(l_j-k_j) J_{1j}\  J_{2j}\  G_{1j}(x)\
    G_{2j}(x)+J_{2j}\ G_{1j}(x)\ H_{2j}(1-x)
\edm
\bdm
   J_{1j}\ J_{2j} (k_j-l_j) = \pi \la_{l_jk_j}
\edm
with
\bdm \la_{l_j k_j}\equiv \cot (l_j\pi )-\cot (k_j\pi ) \; .
\edm

At this point there is a comment in order. It may well happen
that $pk$ $=$ $0$ mod $N$. In this case the system (\ref{clmond})
of linear global monodromy equations given in \cite{bur} is under--determined,
since some of the equations
for $a=2$ are trivially fulfilled. If one would not find additional
independent global monodromy conditions this would imply
a {\em continuous} set of world sheet instantons, which would certainly
be an incorrect result. Therefore one has to look for a modification
of $C_2$.
This situation is not artificial, since it already appears for one of
the simplest $Z_N$ orbifolds:
Consider the two--dimensional $Z_6$ orbifold and choose
$k=2$ and $l=3$ in (\ref{correlation}).
The construction of the loop $C_2$
obviously requires $p=3$ and $q=2$. Hence we have $pk$ $=$ $N$ and
both sides of (\ref{clmond}) for $a=2$ vanish.
Thus there is only one\footnote{The
third relation for $a=3$ merely restricts the set of allowed coset
vectors $v_1$ as will be shown soon.
Hence it cannot help to obtain a unique
solution of (\ref{clmond}) for given $v_1$, $v_2$, $v_3$.}
complex equation
for the two complex coefficients $b$ and $c$.
This problem may be solved
by constructing an alternative loop $C_2$ in the following way:
Starting with $x$ this loop surrounds
the world sheet locations $1$ and $x$
{\em alternately} instead of first encircling $x$ three times and then
twice the point $1$. By multiplying the space group
conjugacy classes of the encircled twist fields in the corresponding
new order one arrives
at the new set of allowed vectors $v_2$:
\bdm F_2 := \fc{(1-\Th^2 )(1-\Th^3 )}{(1-\Th^5 )}
     (f_{23}+\La )\subset \La\; ,
\edm
which is nontrivial (contrary to $F_2$ in (\ref{mv})). Furthermore
when calculating the loop integrals in (\ref{clmond}) carefully,
one finds that the matrix factor $\fc{(1-\Th^2 )(1-\Th^3 )}{(1-\Th^5 )}$
cancels out of (\ref{clmond}) and one recovers the naive result
(\ref{Instcoef}) although the coefficients
$b_2^j$, $c_2^j$ as well as $v_2^j$
now have a different form compared to the singular expressions
(\ref{bbcc}) and (\ref{mv}). In fact the calculation of the
quantum part of (\ref{correlation}) also requires the use of
the new loop $C_2$.

Now we have to find out, which of the solutions (\ref{Instcoef})
are compatible with
(\ref{clmond}) for $a=3$. One easily sees, that the
global monodromy conditions  (\ref{clmond}) for $a=1$ and $a=3$
are consistent with each other, if and only if
\be v_3=-v_1 \; .
\label{minus}
\ee
This relation is either obtained by a
direct evaluation of (\ref{clmond}) for
$a=1,3$. On the other hand it is obvious from figure \ref{C1C2C3}, since
the loops $C_1$ and $C_3$ can be deformed into each other
(up to their relative orientation) by simply
pulling them around the world sheet
sphere $S^2\equiv \CC\cup\{\infty\}$.
The meaning of (\ref{minus}) is, that any instanton is uniquely
determined by two lattice vectors: $v_1 \in F_1 \cap\lf( -F_3\ri)$
and $v_2 \in F_2$. In appendix \ref{sets} it is shown, that
\be \label{sumrange}
F_1 \cap\lf( -F_3\ri) = (1-\Th^k)(f_{21}+\tilde{\La})
\ee
with
\be
\tilde{\La} := -T_1 + \fc{1-\Th^l}{1-\Th^\phi} \La \; .
\label{restricted}
\ee
Here $\phi$ denotes the greatest common divisor of $k$ and $l$ and
$T_1$ is a solution of the space group selection
rule (\ref{sgsr1}) for given $f_i$.

With these preparations we are now able to evaluate
the classical part of (\ref{correlation}) which,
in contrast to the quantum part, depends explicitly on the
antisymmetric background tensor field $B$.
For this purpose we divide
$B$ into $(d/2)^2$\  $2 \times 2$ blocks:
\be
 B_{ij}=-B_{ji}^t=\left( \begin{array}{cl}
 a_{ij} & b_{ij} \\
 c_{ij} & d_{ij}
 \end{array}
 \right)\ \ ,\ \ i,j=1,\cdots,d/2\; ,
 \label{Bblocks}
\ee
such that the block $B_{ij}$ couples the $i$--th with the $j$--th
two--dimensional subplane (which was complexified in (\ref{complex}))
via the term $\ol{\partial}X^t B\partial X$ in (\ref{action1}).

Inserting (\ref{Bblocks}) and  the instanton solutions
(\ref{Ansatz}) together with (\ref{Instcoef}) and (\ref{bbcc})
into the action (\ref{action1}) one encounters various complex
integrals. First of all from the metric part and the $B_{ii}$--blocks
one gets integrals of the type
\be \int d^2z |\omega_{k_i,l_i}(z)|^2 =
 2\pi J_{1i} {\rm Re}\lf[ (1-x)^{k_i-l_i}
 \ol{G}_{2i}(\ol{x})H_{1i}(1-x) \ri]
 +\pi^2 \la_{l_ik_i} |G_{2i}(x)|^2
\ee
and their analogs obtained by
substituting $(k_i,l_i) \rightarrow (1-k_i,1-l_i)$.

The situation is more involved for those parts of the action, which
correspond to blocks $B_{ij}$ with $i\neq j$. In this case one finds
integrands of the type $\omega_{k_i,l_i}(z) \
\bar \omega_{k_j,l_j}(\bar z) $, which are generically not single
valued on the complex plane, and consequently (\ref{action1}) does not
seem to be well defined. We can only give a mathematical meaning
to such integrals if $|k_i|=|k_j|$
(or equivalently $|l_i|=|l_j|$).

What comes to the rescue is the fact, that,
as pointed out in \cite{ejlm}, (\ref{action1})
is only well defined if
\be
 [B,\Theta]\ =\ 0 \; .
 \label{Comm}
\ee
This equation puts strong restrictions on the allowed non--vanishing
components of $B$, which were analyzed in \cite{ejl}. Using the above
decomposition of $B$ into $(\fc{d}{2})^2$ $2\times 2$--blocks, it reduces
to  (no summation over $i$ and $j$)
\bda B_{ij} \vartheta (\al_j) - \vartheta (\al_i) B_{ij} &=& 0 \nnn
     B_{ij} \vartheta^t (\al_j) - \vartheta^t (\al_i) B_{ij} &=& 0 \; .
\eda
It follows that only those blocks $B_{ij}$ are allowed to be
different from zero, that couple two--dimensional subplanes
rotated by the the same or the opposite angle, i.e.
for which  $\cos(2\pi\al_i)$ $=$ $\cos(2\pi\al_j)$.
We have to distinguish three different cases:
\begin{enumerate}
 \item  The two subplanes are rotated by the same angle ($\neq$
 $0$ mod $\pi$) in the same direction, i.e.
 $\sin (2\pi\alpha_i)=\sin(2\pi\alpha_j)\neq 0$. This yields
 \be
  B_{ij}=\left( \begin{array}{rc}
  a_{ij} & b_{ij} \\
  -b_{ij} & a_{ij}
  \end{array}\right) \; .
  \label{case a}
 \ee
 For blocks $B_{ii}$ on the diagonal of course we have $a_{ii}=0$.
 \item  The rotation angles have the same absolute value but opposite
 signs, i.e.  $\sin(2\pi\alpha_i)=-\sin(2\pi\alpha_j)\neq 0$, which
 implies
 \be
  B_{ij}=\left( \begin{array}{cr}
  a_{ij} & b_{ij} \\
  b_{ij} & -a_{ij}
  \end{array}
  \right)\ .
  \label{case b}
 \ee
  This obviously can never happen for blocks $B_{ii}$ on the diagonal.
 \item  Both subplanes are not rotated at all, i.e.
 $\sin(2\pi\alpha_i)=\sin(2\pi\alpha_j) = 0$ and hence
 \be
 B_{ij}=\left( \begin{array}{cl}
 a_{ij} & b_{ij} \\
 c_{ij} & d_{ij}
 \end{array}
 \right)\; .
 \label{case c}
 \ee
  For $i=j$ one finds in addition $a_{ii}=d_{ii}=0$
 and $b_{ii}=-c_{ii}$.
\end{enumerate}

This shows, that due to (\ref{Comm}) indeed
only those integrals contribute
to (\ref{action1}), whose integrands fulfill the condition $|k_i|=|k_j|$
(and equivalently $|l_i|=|l_j|$) and are therefore single--valued on the
complex plane\footnote{Actually the third case is equivalent to fixed
directions of $\Th$, which were excluded before. Nevertheless we took
it into account in order to show, that (\ref{action1}) is well defined
even in this case, provided (\ref{Comm}) holds.}.

It is important to point out here, that the condition (\ref{Comm}) is a
direct consequence of describing the orbifold
CFT with the help of the action
(\ref{action1}). This is however not necessary.
The theory is perfectly well
defined if $B$ fulfills the condition
\bdm [B,\Th ]w \in\La^{\ast}
\edm
for all $w\in\La$ \cite{gso}, although one cannot rely on the action
(\ref{action1}) any longer in this
case \cite{ejlm}. This implies of course as
well, that the so far used path integral measure
does not exist then. Nevertheless it is possible to find a
path integral description of such an orbifold theory with so
called {\em quantized} background fields \cite{Jung}.
This can be done by suitably embedding its Narain--lattice into a
``doubled'' one, whose winding lattice is chosen to be the
Euclidean version of the original Narain--lattice \cite{NSV1,NSV2}
and whose $B$--field commutes with the twist.
This situation is closely related to the
phenomenologically most interesting case of heterotic strings with
quantized Wilson lines \cite{gso}. The techniques outlined in
\cite{Jung} also apply to the calculation of twist field correlations
in such cases.

Collecting all relevant terms and switching back to real quantities we
end up with the following action for an instanton solution
characterized by the two vectors $v$ $\in$ $F_1 \cap (-F_3)$
and $v_2$ $\equiv$ $(1-\Th^{pk})u$ $\in$ $F_2$:
\be
 S_E [G,B,v,u] = \frac{1}{2}v^tCv+\frac{1}{2}u^tDu+\frac{1}{2}u^tXv
 +\frac{1}{2}u^tE^\pr v+i\pi u^tBv+\frac{i\pi}{4}v^tEBv \label{4sv} \; ,
 \label{instaction}
\ee
where we used the following matrices
\be \begin{array}{rcl}
 C &=& {\rm diag}\lf( V_{11,j}\; , 1\leq j\leq\fc{d}{2}\ri)
 \otimes {\bf 1}_2\nnn
 D &=& 4\ {\rm diag }\lf( V_{22,j}\; ,
 1\leq j\leq\fc{d}{2}\ri)\otimes {\bf 1}_2\nnn
 X &=& 4i\ {\rm diag}\lf( \tilde{V}_{12,j}\; ,
 1\leq j\leq\fc{d}{2}\ri)\otimes {\bf 1}_2
 \end{array}
\ee
with\footnote{Note that the exchange $k\leftrightarrow l$ acts as
$J_{1j}\leftrightarrow J_{2j}$, $H_{1j}\leftrightarrow H_{2j}$ and
$G_{1j}\leftrightarrow G_{2j}$.}
\bea
 V_{11,j} &=& \ds{
 \lf( 2 |I_j (x,\ol{x})| \ri)^{-2}
 \Big [ J_{2j}^2|H_{2j}(1-x)|^2 \Big ( 2J_{1j}
 {\rm Re}\lf\{ G_{1j}(x)\bar H_{1j}(1-\ol{x}) \ri\} } \nnn
 && \hspace{3.2cm} + \ds{\pi \la_{l_j k_j} |G_{1j}(x)|^2 \Big )
 + (k\leftrightarrow l)\Big ] }\nnn

 V_{22,j} &=& \ds{ \frac{\pi^2}{2|I_j (x,\ol{x})|^2}
 {\rm Re}\lf[ G_{1j}(x)G_{2j}(x)
 \lf(J_{1j} \bar G_{2j}(\ol{x}) \bar H_{1j}(1-\ol{x})
 + (k\leftrightarrow l)\ri) \ri] }\nnn

 \tilde{V}_{12,j} &=& \ds{ \frac{i\pi J_{1j}J_{2j}}
 {2 |I_j (x,\ol{x})|^2}
 {\rm Im}\lf[ G_{1j}(x)G_{2j}(x)
 \bar H_{1j}(1-x) \bar H_{2j}(1-x) \ri] } \; .
\eea
The block--diagonal and anti--symmetric matrices $E$ and $E^\pr$ are
defined as
\bdm
 \fc{1}{2}E = \frac{1}{1-\Theta^k}-\frac{1}{1-\Theta^l}
\edm
\bdm
 E^\pr_{2j-1,2j} = -E^\pr_{2j,2j-1} = 4A_j \; ,
\edm
 where
\bdm A_j = \fc{\pi}{4 |I_j (x,\ol{x})|^2}
     \lf[ J_{2j}^2|G_{1j}(x)|^2|H_{2j}(1-x)|^2
     - (k\leftrightarrow l)\ri] \; .
\edm
They satisfy the useful identity
\bdm 4CD=-E^{\prime 2}+X^2+\pi^2{\bf 1}\; .
\edm

The quantum part $Z_{qu}(x,\bar x)\equiv \int\Cd X_{qu}
\exp\{-S_E [G,B,X_{qu}]\}$ of (\ref{correlation})
may be determined by the stress energy
tensor method \cite{dix,bur}. Since
the $B$--term in (\ref{action1}) is a total derivative it only
yields non--vanishing contributions for topologically nontrivial
instanton solutions, which do not affect $Z_{qu}$.
We may therefore simply use the result obtained in \cite{bur} for
$B=0$ which is given by
\bdm Z_{qu}(x,\xb ) = \nu\ |x|^{-4h_k^\si}
     \lf[\prod_{j=1}^{d/2}\fc{(1-x)^{-l_j (1-k_j )}
     (1-\xb )^{-k_j (1-l_j )}}{I_j (x,\xb )}\ri]  \; .
\edm
The integration constant $\nu$ will be fixed in section \ref{YY}
by conventionally normalizing the  two twist twist field correlation
function according to (\ref{norm}). Hence we end up with
\bea Z_{\{f_{i}\}}(x,\bar x) &=& \ds{\nu\ |x|^{-4h_k^\si}
     \lf[\prod_{j=1}^{d/2}\fc{(1-x)^{-l_j (1-k_j )}
     (1-\xb )^{-k_j (1-l_j )}}{I_j (x,\xb )}\ri] } \nnn
     &\times& \ds{\sum_{{v\in F_1\cap \ (-F_3)}\atop
     {u\in f_{23} +\La}}
     e^{-S_E [G,B,v,u]}} \; .
     \label{FTClagr}
\eea

\section{S--channel factorization}
\label{YY}

Space group considerations and $SL(2,\CC )$--invariance tell us, that
the OPE between a twist and an anti--twist
field must be of the form\footnote{
The factor $\fc{1}{N}$ is a convention and
serves to undo an $N$--fold overcounting due to
$V_{p,w}^{\rm inv}=V_{\Th p,\Th w}^{\rm inv}$ if $(p,w)\ne(0,0)$.
Notice that $V_{0,0}^\inv = \sq{N}{\bf 1}$ differs from the ordinary
identity operator.}
\be
   \sigma^{-k}_{f_a}(0,0)\sigma^{k}_{f_b}(x,\bar{x})
   =\frac{1}{N} \hspace{-.25cm}\sum_{p\in\Lambda^{\ast}\atop
   w\in(1-\Th^k)(f_{ba}+\La)} \hspace{-.2cm}x^{h-2h_k^\si}\
   \bar{x}^{\bar{h}-2h_k^\si}\ C^k_{f_b,f_a;p,w}\
   V_{p,w}^{\rm inv}(x,\bar{x})\ + \ldots \ ,
   \label{ope1}
\ee
where $V_{p,w}^{\rm inv}$ denotes a twist invariant vertex operator
of the untwisted sector, describing the creation of
an untwisted string state with momentum $p\in\La^{\ast}$
and winding $w\in\La$:\footnote{Of course,
in the case when $V_{p,w}^{\rm inv}$
is really an operator, i.e.\ in a canonically quantized theory, normal
ordering of the exponentials is required.}
\be
   V_{p,w}^{\rm inv}=\fc{1}{\sqrt{N}}\sum_{k=0}^{N-1}
   e^{i (\Th^k P_L)^t X_L(z)
   +i (\Th^k P_R)^t X_R(\bar{z})}\ =\fc{1}{\sqrt{N}}\sum_{k=0}^{N-1}
   V_{\Th^k p,\Th^k w}(z,\bar{z})
   \label{vertexoperator}
\ee
with
\bda
    P_R &\equiv& \ds{p+(G-B)w = p+\frac{1}{2}w-Bw} \nnn
    P_L &\equiv& \ds{p-(G+B)w = p-\frac{1}{2}w-Bw}
\eda
and conformal weights
\bda
   h &\equiv& \ds{\frac{1}{4}P_R^t G^{-1} P_R
   = \frac{1}{2} P_R^t P_R} \nnn
   \bar h &\equiv& \ds{\frac{1}{4} P_L^t G^{-1} P_L
   = \frac{1}{2} P_L^t P_L} \; .
\eda
Using the definition
\be C^l_{f_4,f_3;p,w}:=\lim_{|z_{\iy}|\ra \iy} |z_{\iy}|^{4h_l^\si} \,
    \lng V_{-p,-w}^{\rm inv}(0,0)\si_{f_3}^{-l}(1,1)
    \si_{f_4}^{l}(z_{\iy},\zb_{\iy}) \rng \label{emi}
\ee
of the OPE coefficient and (\ref{ope1}), it
is clear, that $Z_{\{f_{i}\}}(x,\bar x)$ must behave as
\be Z_{\{f_{i}\}}(x,\bar x) = \frac{1}{N}\
     \sum_{p \in \La^{\ast} \atop v \in F_1
     \cap (-F_3)}
     C_{f_2,f_1;p,v}^k\ C_{f_4,f_3;-p,-v}^l|x|^{-4 h_k^\si} x^h\
     \bar x^{\bar h} +\ldots  \label{limit00}
\ee
for $x,\xb\ra\ 0$.
The summation range of the allowed winding vectors $v$ is fixed by
requiring that {\em both} three--point functions are allowed to be
non--vanishing in view of their space group selection rule.
Already this fact clearly
indicates, that the intersection of $F_1$ with $-F_3$ is required
instead of $F_1$ alone as in \cite{bur}.

The actual behaviour of $Z_{\{f_{i}\}}$ for
$|x|\ra 0$ is obtained by first
performing a Poisson resummation of (\ref{FTClagr})
in $u^\prime$ $:=$ $u$ $-$ $f_{23}$ \ \ \
and then inserting the asymptotic form of the hypergeometric functions
\cite{as}. This yields (for details see \cite{stieber})
\bea
    \ds{Z_{\{f_i\}}(x,\bar x)} &\approx& \ds{\nu\
    \fc{|x|^{-4 h_k^\si}}{(2\pi )^{d/2}V_{\La}}} \nnn
    &\times& \ds{\sum_{p \in \La^{\ast} \atop v \in F_1
     \cap (-F_3)} \hspace{-.5cm}
    e^{-2 \pi if_{23}^t p
    + i\pi p^t\lf(\fc{1}{1-\Th^k}-\fc{1}{1-\Th^l}\ri)v}
    \lf[ \prod_{j=1}^{d/2}\ \frac{x^{h^j} \ \ \bar x^{\bar h^j}}
    {\dt^{\fc{1}{2}(h^j+\bar h^j)}_{k_j}
    \ \dt^{\fc{1}{2}(h^j+\bar h^j)}_{l_j}} \ri] }
    \label{limit0}
\eea
where
\bdm\ba{rclcrcl}
    h^j &\equiv& \ds{\frac{1}{2} \sum\lm_{r=2j-1}^{2j}
    \lf[(p+\frac{1}{2}v-Bv)^r\ri]^2} &,&
    \bar h^j &\equiv& \ds{\hal \sum\lm_{r=2j-1}^{2j}
    \lf[ (p-\frac{1}{2}v+Bv)^r \ri]^2} \nnn
    \ln \dt_{k_j} &\equiv&2 \psi(1)-\psi(k_j)-\psi(1-k_j) &,&
    \psi(x) &\equiv& \ds{\fc{d \ln\Gamma(x)}{dx}}\; ,  \ea
\edm
and $V_\La$ denotes the Volume of the unit cell of $\La$, i.e.\
$V_\La =\sqrt{\det g}$.

The normalization constant $\nu$ will now be fixed according
to (\ref{norm}).
This corresponds to the choice
\be  C_{f,-f;0,0}^k =\sq{N} \;\Ra\; \nu
     = V_\La (2\pi )^{d/2}\; .
     \label{const}
\ee
Finally by comparing (\ref{limit0})
with (\ref{limit00}) we find
\bdm
   C_{f_b,f_a;p,w}^k=\sqrt{N}\
   \lf[ \prod_{j=1}^{d/2}\dt^{-\hal (h^j+\bar h^j)}_{k_j}\ri] \ \
   e^{-2\pi i f_b^t p}\ \ e^{i \pi p^t \fc{1}{1-\Th^k}w}\ \
   \dt_{w\in (1-\Th^k)(f_{ba}+\La )} \; ,
\edm
which agrees perfectly with the results obtained in \cite{ejlm} by
operator methods.

Note that this three--point coupling could only be inferred from
(\ref{FTClagr}), since the restricted summation range $F_1$
$\cap$ $(-F_3)$ was used instead of $F_1$ as in \cite{bur}.
Hence we conclude, that it is mandatory to take the
global monodromy loop $C_3$ into account in order to
obtain a fully crossing symmetric\footnote{
The factorization of (\ref{FTClagr}) into the twisted $u$--channel
will be proven in the next section.}
four--point function (\ref{FTClagr}).

Using the identity \cite{ejlm}
\bea
    g_k(P_L,P_R) &\equiv& \ds{\sqrt{N}\ \prod_{i=1}^{d/2}
    \dt_{k_i}^{-\frac{1}{2}(h^i+\bar h^i)}} \nnn
    &=& \ds{\sqrt{N}\ \lf[ \frac{1}{N} \ri]^{\frac{1}{2}(P_R^2+P_L^2)}
    \prod_{n=1}^{N-1}|1-e^{2\pi in/N}|^{\fc{1}{2}(P_R^t
    \Theta^{kn} P_R+P_L^t \Theta^{kn} P_L)}}
    \label{gprime}
\eea
we obtain the final result
\be
   C_{f_b,f_a;p,w}^k =
   g_k (P_L,P_R)\   e^{-2\pi i f_b^t p}\ \
   e^{i \pi p^t \frac{1}{1-\Theta^k}w}\ \
   \delta_{w\in (1-\Th^k)(f_{ba}+\Lambda)}\ .
   \label{ssv}
\ee
This expression will be especially useful for the discussion of discrete
background symmetries in section \ref{sec6},
since it avoids the splitting of
the target space into $d/2$ two--dimensional subspaces.
It can be simplified considerably for orbifolds with prime $N$,
which allows for an easy proof of duality invariance in these cases
\cite{stieber}.

\sect{U--channel factorization}
\label{YYY}

The Yukawa couplings can be found by factorizing the four point function
$Z_{\{f_i\}}(x,\bar x)$ into the $k+l$--twisted or u--channel, that is
by considering the limit $|x|\ra\infty$. The resulting
expression can be interpreted as
a sum of terms bilinear in the Yukawa couplings by using the OPE
\be
    \si_{f_2}^{+k}(x,\bar x) \si_{f_4}^{+l}(z_{\infty},\bar z_{\infty})
    =  |x-z_{\infty}|^{2(h_{k+l}^\si -h_k^\si -h_l^\si )}
    \sum_{f \in \wp_{24}} Y_{f_2,f_4,f}^{k,l}\
    \si_{f}^{+(k+l)}(z_{\infty},\bar z_{\infty})+ \ldots
    \label{ope} \ .
\ee
In this equation $\wp_{24}$ denotes the set of
all fixed points $f$ for which the space group selection rule
allows a non--vanishing three point correlator for given $f_2$ and $f_4$.
Thus $\wp_{24}$ is the set of all $f$ for which there exist
$\tau_1$, $\tau_2$ $\in\La$ such that
\be
   (1-\Th^k)f_2+\Th^k(1-\Th^l)f_4
   -(1-\Th^{k+l})f=(1-\Th^k)\tau_1+\Th^k(1-\Th^l)\tau_2 \; .
   \label{sg}
\ee

The operator product coefficients
\bdm Y_{f_a,f_b,f_c}^{k,l}\equiv
 \lim_{|x| \ra \infty} |x|^{4h_k^\si}
 \lng \si_{f_a}^k(x,\bar x)\si_{f_b}^l(1,1)
     \si_{f_c}^{-(k+l)}(0,0)\rng
\edm
may be determined by considering the asymptotic behaviour of
$Z_{\{f_i\}}(x,\bar x)$ for $|x|\ra\infty$:
\be Z_{\{f_i\}}(x,\bar x)=
    |x|^{2(h_l^\si -h_k^\si -h_{k+l}^\si )}\sum_{f \in \wp_{24}}
    Y_{f_1,f_3,f}^{\ast k,l} Y^{k,l}_{f_2,f_4,f} + \ldots \ .
\label{cftyy}
\ee

At this point we will extend the assumption of
section \ref{fourpointcorrelation} that $\Th^k$ and $\Th^l$ do not have
fixed directions to $\det (1-\Th^{k+l})\neq 0$. If we would allow for
invariant subplanes of $\Th^{k+l}$ similar arguments as given in
section \ref{fourpointcorrelation} would apply.
The corresponding part of the
Yukawa coupling $Y_{f_a,f_b,f_c}^{k,l}$ would
actually again be a two--point
function of twist fields as in (\ref{norm}).
This is however, as mentioned
already, not the whole story because of
possible momenta and windings in the
invariant directions. They would allow for a full tower of additional
three--point
functions of twisted sector fields. We will defer the discussion of these
issues to section \ref{fixtori}.

For the factorization (\ref{cftyy}) to take place in (\ref{FTClagr}),
the instanton action (\ref{instaction}) has to decompose into two
summands, one of which depends only on the
fixed points $f_1$ and $f_3$, and the other only on $f_2$
and $f_4$. This is achieved by introducing the new
coset vectors
\bea \ds{\tilde{v}_1} &:=&
     \ds{-v_2+\fc{1-\Th^{pk}}{1-\Th^k} v_1} \nnn
     \ds{\tilde{v}_2} &:=&
     \ds{v_2-\fc{1-\Th^{pk}}{1-\Th^l}v_3\ .}
\label{tv1tv2}
\eea
This substitution
can be understood as choosing closed net twist zero loops
$\tilde C_1$, $\tilde C_2$ in such a way that
they encircle $(z_1,z_3)$, $(z_2,z_4)$
rather than $(z_1,z_2)$, $(z_2,z_3)$. The corresponding coset
vectors coincide with the ones given above.

To extract the leading order term of (\ref{FTClagr}) for $|x|\ra\infty$,
we use the asymptotic behaviour of the hypergeometric functions
in this limit \cite{as}. One finds that
the asymptotic form of the classical
action \req{instaction} reads
\bda S_E &=& \ds{2\pi \Big[
     \tilde{v}_1^t\fc{1}{\Id -\Th^{-pk}}(\fc{\bf 1}{2}-B)
     L\fc{1}{\Id -\Th^{pk}}\tilde{v}_1} \nnn
     &+& \hspace{.66cm}\ds{\tilde{v}_2^t\fc{1}{\Id -\Th^{-pk}}
     (\fc{\bf 1}{2}+B)L\fc{1}
     {\Id -\Th^{pk}}\tilde{v}_2 \Big] } +
     \ldots \ ,
\eda
where we defined the anti--symmetric and block--diagonal matrix
\be
\ba{clccccr}
    L_j &=&  {\ds{1}\ov\ds{|\lambda_j|}}
   \lf( \Eb_2+ i\,{\rm sgn}(\lambda_j\ri) \epsilon) &,& \epsilon
    = \lf( \ba{clcr} 0 & 1 \nnn -1 & 0 \ea \ri) \, ,  \nnn
    \ds{\la_j} &:=&
    \ds{\cot(\pi k_j)+\cot(\pi l_j)}
    &,& \ds{j=1,\ldots,d/2\ .}
\ea
\ee
This way one arrives at the following asymptotic form of the four point
correlation function:
\bea
     \ds{Z_{\{f_i\}}(x,\bar x)} &{\bf \approx}& (2\pi)^{\fc{d}{2}}
     V_{\Lambda}\ds{|x|^{2
     \sum \limits_{i=1}^{d/2}(k_i+l_i-1)k_i} \lf[\prod_{j=1}^{d/2}
     \Ga^2_{k_j,l_j}\ri] } \nnn
     &\times& \ds{\sum_{{w_1 \in
     \cu_1} \atop {w_2 \in \cu_2}} e^{-2\pi[w_1^t({\bf \hal}-B)L
     w_1+w_2^t({\bf \hal}+B)L w_2]}\ +\ldots\ ,}
\eea
where $w_i \equiv (1-\Th^{pk})^{-1} v_i$, and
\be \ba{cll}
    \ds{\Ga_{k_j,l_j}} &=& \left\{ \ba{c@{\quad:\quad}l}
    {\ds{\Ga(1-k_j)\Ga(1-l_j)\Ga(k_j+l_j)}
    \ov\ds{\Ga(k_j)\Ga(l_j)\Ga(1-k_j-l_j)}} &
    \ds{0<k_j+l_j<1} \\
    \noalign{\vspace{0.3cm}}
    \ds{\Ga_{1-k_j,1-l_j}} & \ds{1<k_j+l_j<2\ .}\ea\right.
    \ea \label{abbrev}
\ee
The summation ranges over the sets
\bea
\cu_1 &=& \ds{\{ f_{31} - T_1 - \lambda + \frac{1-\Th^l}{1-\Th^\phi} \mu \mid
\lambda, \mu \in \Lambda \} } \nnn
\cu_2 &=& \ds{\{ f_{24} + T_2 + \lambda + \frac{1-\Th^k}{1-\Th^\phi} \mu \mid
\lambda, \mu \in \Lambda \} }\; .
\eea

While the summands are
factorized, the sums do not actually decouple
as in (\ref{cftyy}). This is
because via (\ref{sgsr1}) $T_1$ and $T_2$ appearing here
depend on both $f_1,f_3$ and $f_2,f_4$.
To achieve the required decoupling we proceed in a way, which
generalizes the method used in \cite{rp} for the case $k=l =1$. This
entails
changing summation variables appropriately.
Observe first that the summation
range $\cu_1$ can be written in the form
\eq
    \cu_1 = \{f_{31} + \tau + \frac{1-\Th^{k+l}}{1-\Th^\phi} \alpha \mid
    \alpha \in \Lambda, \tau \in
    \frac{\Lambda}{\frac{1-\Th^{k+l}}{1-\Th^\phi} \Lambda} \}\; .
\qe
The reason for writing it  in  this way is that the
quotient space over which $\tau$
runs can be represented in terms of the set
of all $\tau_1 - \tau_2$, which
appear in the three point selection rule:
\be
   (1-\Th^k)f_1+\Th^k(1-\Th^l)f_3
   -(1-\Th^{k+l})f=(1-\Th^k)\tau_1+\Th^k(1-\Th^l)\tau_2 \; ,\label{sg13}
\ee
The set of fixed points $f$ satisfying (\ref{sg13})
for some lattice vectors
$\tau_1$ and $\tau_2$ is denoted by $\wp_{13}$ and is
equal to $\wp_{24}$.

The summation range can now be written in the form
\eq
\cu_1 = \bigcup_{f \in \wp_{13} } \cu_1(f),
\qe
where
\eqr
 \cu_1(f) = &\{ & \hspace{-.3cm} f_{31} + \tau_1 - \tau_2 +
\frac{1-\Th^{k+l}}{1-\Th^\phi} \alpha \mid  \alpha \in \Lambda, \\
\nonumber
& & (1-\Th^k)(f_1-\tau_1)+\Th^k(1-\Th^l)(f_3-\tau_2)
=(1-\Th^{k+l})f \} \ .
\rqe
This follows from the fact that for a given $f$ $\in$
$\wp_{13}$, the set of $\tau_1, \tau_2$
which solve \req{sg13} is such, that
the difference $\tau_1 - \tau_2$ is fixed modulo
$\fc{1-\Th^{k+l}}{1-\Th^\phi} \Lambda\ .$

The argument given above is a prescription for changing the summation
variables from $\lambda, \mu$ to $\alpha, f$. After this change is made,
the summation range for $w_1$ will depend only on the fixed points
$f_1, f_3$. One may check, that
the above change of variables leads to the following summation range for
$w_2$:
\eq
\cu_2 = \bigcup_{f \in \wp_{24} } \cu_2(f),
\qe
where
\eqr
\cu_2(f) = &\{ & \hspace{-.3cm} f_{24} -\rho_1 +\rho_2 +
\frac{1-\Th^{k+l}}{1-\Th^\phi} \alpha \mid  \alpha \in \Lambda, \\
\nonumber
& & (1-\Th^k)(f_2-\rho_1)+\Th^k(1-\Th^l)(f_4-\rho_2)
=(1-\Th^{k+l})f \} \ .
\rqe
This depends only on $f_2, f_4$, so the limiting form of the four point
correlation function can now be written as
\bdm \ds{Z_{\{f_i\}}(x,\bar x)
     {\ \approx\ } (2\pi)^{\fc{d}{2}} V_{\Lambda}
     |x|^{2(h_l^\si-h_k^\si-h_{k+l}^\si)}
     \lf[\prod_{j=1}^{d/2} \Ga^2_{k_j,l_j}\ri]}
\edm
\be \ds{\times \sum_{f \in \wp_{24}=\wp_{13}}\sum_{w_1 \in {\cu_1(f)}}
    e^{-2\pi w_1^t({\bf \hal}-B)L w_1}\ \sum_{w_2 \in {\cu_2(f)}}
    e^{-2\pi w_2^t({\bf \hal}+B)L w_2}\ +\ldots\ .}
\ee
This can be compared directly  with (\ref{cftyy})
and one finds
\be
\ds{Y_{f_a,f_b,f_c}^{k,l} = (2\pi)^{\fc{d}{4}}
\lf[ V_\La
\prod_{j=1}^{d/2} \Ga_{k_j,l_j}\ri]^{1\ov2}
\sum_{v \in {\cal U}}  e^{-2\pi v^t({\bf \hal}+B)Lv}\ ,}
\label{sss}
\ee
where
\be {\cal U} = \ds{f_{ba} - \tau_{ba}
    + {1-\Th^{k+l}\ovx 1-\Th^\phi}\La\; ,}
\ee
and $\tau_a, \tau_b$ are defined by the three point selection rule for
$f_a, f_b$ analogously to (\ref{sg13}).

In appendix \ref{direct} it will be shown that this result is consistent
with the global monodromy conditions for the three--point coupling.

\sect{Fixed tori}
\label{fixtori}

We now intend to analyze the situation,
where the orbifold possesses fixed tori,
i.e.\ $\det (\Id -\Th )=0$ \cite{Jung}.
The generalization of the following
considerations to higher twisted sectors ($\det (\Id -\Th^k )=0$, $k>1$)
is straightforward. Part of this discussion may be found in
\cite{NSV1,NSV2}.

We first apply a coordinate transformation such that the generating twist
matrix is block--diagonal:
\bdm \Th =\lf( \ba{cc} \vartheta & 0 \\ 0 & \Id_{d_2} \ea \ri) \; ,
\edm
where $\vartheta\in O(d_1)$, $\det (\Id_{d_1}-\vartheta )\neq 0$
and $d_1+d_2=d$.
W.r.t.\ these coordinates the antisymmetric tensor field takes on the
form
\be B=\lf( \ba{cc} B_1 & 0 \\ 0 & B_2 \ea \ri) \; .
    \label{B1B2}
\ee
Possible non--vanishing off--diagonal blocks which would couple
twisted and invariant directions are excluded by the requirement
$[B,\Th ]$ $=$ $0$ \cite{ejlm}.

It proves to be useful to define the following sublattices of $\La$:
\bda I&:=&\lf\{ w\in\La\, |\, \Th w=w\ri\} \nnn
     J&:=&\lf\{ v\in\La\, |\, v\cdot I=0\ri\}\; .
\eda
Hence $I$ is the sublattice of $\La$, which is invariant under $\Th$
and $J$ is its orthogonal complement in $\La$. If we denote the
projectors onto the invariant and non--invariant directions of
$\La$ by $\Pc$ and $\Pc^\perp$, respectively, then
\bdm \til{I}:=\Pc \La\;\; ,\;\; \til{J}:=\Pc^\perp\La \; .
\edm
Note that in general $\til{I}$ and $\til{J}$ are not sublattices of
$\La$ and are therefore different from $I$ and $J$.
However we always have
$I\subset\tilde{I}$ and $J\subset\tilde{J}$. Equivalently we find
$\til{I}^{\ast}\neq I^{\ast}$,
$\til{J}^{\ast}\neq J^{\ast}$, $I^{\ast}\nsub\La^\ast$
and $J^{\ast}\nsub\La^{\ast}$. But realize, that for these dual lattices,
marked by asterisks,
the relations $\tilde{I}^{\ast}\subset I^{\ast}$ and
$\tilde{J}^{\ast}\subset J^{\ast}$ hold.

It may be helpful to illustrate these definitions
with a simple example. For this purpose we choose a two--dimensional
$Z_2$ orbifold, whose winding lattice is the root lattice
of the Lie algebra $su(3)$. The  twist is assumed to be a reflection
w.r.t.\ the $x^2$--axis. We obtain
\bda \La&=&\lf\{n^1e_1+n^2e_2\, |\, n^1,n^2\in\ZZ ,\;\;
     e_1=\lf( \ba{c} \sqrt{2} \\ 0 \ea \ri) ,\;\;
     e_2=\lf( \ba{r} -\sqrt{\fc{1}{2}} \\
     \sqrt{\fc{3}{2}} \ea \ri) \ri\} \nnn
     \La^{\ast}&=& \lf\{ m_1e^{1{\ast}}
     + m_2e^{2{\ast}}\, |\, m_1,m_2\in\ZZ ,\;\;
     e^{1{\ast}}=\lf( \ba{c} \sqrt{\fc{1}{2}} \\
     \sqrt{\fc{1}{6}} \ea \ri) ,\;\;
     e^{2{\ast}}=\lf( \ba{c} 0 \\ \sqrt{\fc{2}{3}} \ea \ri) \ri\} \nnn
     I&=& \lf\{ n\lf( e_1 + 2e_2 \ri) \, |\,
     n\in\ZZ \ri\} \nnn
     J &=& \lf\{ ne_1 \, |\,
     n\in\ZZ \ri\}    \nnn
     \til{I} &=& \lf\{ n\lf( \ba{c} 0 \\ \sqrt{\fc{3}{2}} \ea \ri)\, |\,
     n\in\ZZ \ri\} \nnn
     \til{J} &=& \lf\{ n\lf( \ba{c} \sqrt{\fc{1}{2}} \\ 0 \ea \ri)\, |\,
     n\in\ZZ \ri\} \; .
\eda

With these examples in hand it is easy to visualize that
\bda \La\sub\tilde{I}\oplus\tilde{J}
     &{\rm but \;\; in \;\; general}&
     \La\neq\tilde{I}\oplus\tilde{J} \nnn
     \La^{\ast}\sub I^{\ast}\oplus J^{\ast}
     &{\rm but \;\; in \;\; general}&
     \La^{\ast}\neq I^{\ast}\oplus J^{\ast} \; .
\eda

The one--loop partition function of this model is obtained via standard
techniques. For the untwisted sector one finds
\bea \Zc_U (\tau ,\ol{\tau})
     &=&\ds{\hal\fc{|e^{2\pi i\tau}|^{-\fc{1}{12}}}
     {\lf|\prod\lm_{n=1}^{\infty}
     (1-e^{2\pi i\tau n})\ri|^{2}}
     \sum_{v\in\La\atop q\in\La^{\ast}}e^{\pi i\tau Q_R^2}
     e^{-\pi i\ol{\tau}Q_L^2}} \nnn
     &+&\ds{\hal\fc{|e^{2\pi i\tau}|^{-\fc{1}{12}}}
     {\lf|\prod\lm_{n=1}^{\infty}
     (1+e^{2\pi i\tau n})\ri|^{2}}
     \sum_{v\in I \atop q\in\til{I}^{\ast}}
     e^{\pi i\tau Q_R^2}e^{-\pi i\ol{\tau} Q_L^2}} \label{ZcU}
\eea
with $Q_{R/L}=q-B v\pm\hal v$.
Performing modular transformations the partition function for the single
twisted sector turns out to be
\bea \Zc_T (\tau ,\ol{\tau}) &=&\ds{
     \hal \fc{|e^{2\pi i\tau}|^{\fc{1}{24}}}{\lf| \prod\lm_{n=1}^{\infty}
     (1-e^{2\pi i\tau (n-\hal )})\ri|^2}
     \sum_{v\in\til{I} \atop q\in I^{\ast}}
     e^{\pi i\tau Q_R^2}e^{-\pi i\ol{\tau}Q_L^2}}\nnn
     &+&\ds{
     \hal \fc{|e^{2\pi i\tau} |^{\fc{1}{24}}}{\lf|
     \prod\lm_{n=1}^{\infty}
     (1+e^{2\pi i\tau (n-\hal )})\ri|^2}
     \sum_{v\in\til{I} \atop q\in I^{\ast}}
     e^{2\pi i q^tv}e^{\pi i\tau Q_R^2}e^{-\pi i\ol{\tau} Q_L^2}}  \; .
     \label{ZcT}
\eea

Starting from these expressions and their generalizations
for arbitrary $Z_N$ orbifolds with fixed tori, we are able to
extract the spectrum of such theories. Coordinate fields corresponding to
twist--invariant directions of $\Th$ will be denoted by $Y$ and those
belonging to the twisted ones by $X$.
The primary fields of the
untwisted sector are now given by the twist invariant
vertex operators (compare (\ref{vertexoperator}))
\bdm V_\Pb^{\rm inv}(z,\zb ) = E_Y(Q_R,Q_L;z,\zb )
     \fc{1}{\sqrt{N}}\sum^N_{j=1}
     E_X(\vartheta^j T_R,\vartheta^j T_L;z,\zb) \; ,
\edm
where
\bda E_Y(Q_R,Q_L;z,\zb ) &=&
     \exp\lf\{ Q_R^t Y_R(z) + Q_L^t Y_L(\zb )\ri\} \nnn
     E_X(T_R,T_L;z,\zb ) &=&
     \exp\lf\{ T_R^t X_R(z) + T_L^t X_L(\zb )\ri\} \; .
\eda
The Narain momentum $\Pb$ has been split into its untwisted and
twisted part:
\bda \Pb &\equiv& (P_R,P_L)\equiv (T_R,Q_R;T_L,Q_L)\nnn
     &=&(p-B_1w+\hal w,q-B_2v+\hal v;p-B_1w-\hal w,q-B_2v-\hal v)
\eda
with
$w\in\tilde{J}$, $v\in\tilde{I}$ such that $w+v\in\La$ and
$p\in J^{\ast}$, $q\in I^{\ast}$ such that $q+p\in\La^{\ast}$.

Note, that compared to the path integral quantization
performed in this paper,
an operator quantization of the orbifold theory requires
a different expression for twist invariant vertex operators
of the untwisted sector. Their $l$--th twisted sector
representations take on the form
\bea V_\Pb^{\rm inv}(z,\zb ) &=& :E_Y(Q_R,Q_L;z,\zb ): \nnn
     &\times& \ds{\fc{1}{\sqrt{N}}\sum^N_{k=1}
     g_l (\vartheta^k T_R,\vartheta^k T_L)z^{-\fc{1}{2}T_R^2}
     \zb^{-\fc{1}{2}T_L^2}
     :E_X(\vartheta^k T_R,\vartheta^k T_L;z,\zb): } \; ,
     \label{tivertex}
\eea
where the   twisted sector string coupling constant $g_l$
is given by (\ref{gprime}).
Furthermore twisted and untwisted sector mode expansions have to be
inserted for the fields $X_{R/L}$ and $Y_{R/L}$, respectively
and the colon denotes normal ordering.
Appropriate commutation relations of the fields $V_\Pb^{\rm inv}$ are
achieved either by introducing additional cocycle operators or
by carefully quantizing the associated zero modes \cite{ejlm}.

The partition function (\ref{ZcT})
shows, that all states belonging to
twisted sectors with fixed tori possess
completely fixed (i.e.\ not only fixed up to coset shifts) winding
quantum numbers $v\in\til{I}$ w.r.t.\ the invariant directions.
Hence the (non--excited) twist fields have the general form
\be {\cal T}^+_{f_{\tilde{J}},Q}(z,\zb ) :=
    \si_{f_{\tilde{J}}}^+(z,\zb )E_Y(Q_R,Q_L;z,\zb )\; , \label{sigmaqv}
\ee
where $Q_{R/L}=q-B_2v\pm\hal v$, $w \equiv (1-\vartheta )f_{\tilde{J}}$
$\in$ $\til{J}$ and $v\in\til{I}$ such that
\be \label{wv} \begin{array}{c}
    w + v \in \La \\
    q^t v \in \ZZ \; .
\end{array} \ee

Note that the fields (\ref{sigmaqv}) now
generally have different conformal weights for different $f_{\tilde{J}}$,
contrary to the degenerate case without fixed tori.
This is due to the fact, that for given fixed point windings
$w=(1-\vartheta )f_{\tilde{J}}\in\til{J}$ the allowed
windings $v$ $\in$ $\tilde{I}$ of the exponential
vertex operators $E_Y(Q_R,Q_L;z,\zb )$ are restricted by (\ref{wv}).

We will now show that the origin of the condition $q^tv$ $\in$ $\ZZ$
may not only be read off from the twisted sector partition
function (\ref{ZcT}) but can alternatively
be deduced as the requirement of twist invariance for the
fields (\ref{sigmaqv}). We first associate to any such field
a corresponding state $|w,v,q\rng$ of the underlying
two--dimensional CFT. Since
the orbifold is {\em toroidal}, shifting any quantity of the theory
by $2\pi u$ with $u$ $\in$ $\La$ should not affect the theory.
Hence we especially have to identify the states
$|w,v,q\rng$ and $T_{2\pi u}|w,v,q\rng$, where $T_{2\pi u}$ $=$
$e^{ -2\pi i u_{\tilde{J}}^t \hat{p} }$
$e^{ -2\pi i u_{\tilde{I}}^t \hat{q} }$
is the unitary lattice translation
operator on the Hilbert space of states,
whenever $u$ $\equiv$ $u_{\tilde{J}}$ $+$ $u_{\tilde{I}}$ $\in$ $\La$.
$\hat{q}$ and $\hat{p}$ denote the momentum operators w.r.t.\
the invariant and twisted directions, respectively. This condition
implies
\be |w,v,q\rng \equiv e^{-2\pi iu^t_{\tilde{I}}q}
    |w+u_{\tilde{J}},v,q\rng \; .
    \label{cosetstab}
\ee

After this preparation we can investigate the requirement of
twist invariance for the fields (\ref{sigmaqv})
under a space group element $g=(\Th , u)$.
This means
\bda {\cal T}^+_{f_{\tilde{J}},Q}(z,\zb )
     &\stackrel{!}{=}&
     {\cal T}^+_{f_{\tilde{J}},Q}(e^{2\pi i}z,e^{-2\pi i}\zb ) \nnn
     &=& \si^+_{g^{-1}f_{\tilde{J}}}(z,\zb )
     e^{i\{ Q^t_R(Y_R(z)+2\pi u_R^{\tilde{I}})+
            Q^t_L(Y_L(\zb )+2\pi u_L^{\tilde{I}}) \} }
\eda
with $u_{R/L}^{\tilde{I}}:=\hal u_{\tilde{I}}\mp B_2 u_{\tilde{I}}$
(cf. \cite{ejlm}). Using (\ref{cosetstab})
we easily find that this condition
indeed reduces to $q^tv$ $\in$ $\ZZ$.

After this general treatment of some properties
of the spectrum of orbifold
theories with fixed tori we will now consider
correlation functions. Any vertex operator of such theories
obviously factorizes into two pieces (see e.g.\ (\ref{tivertex})
and (\ref{sigmaqv})), one
depending only on the twisted directions and the other on the untwisted
ones: $O [X,Y]$ $=$ $O[X] O[Y]$.
Due to the fact that the $B$--field in (\ref{B1B2}) is block--diagonal,
furthermore also the action (\ref{action1}) decomposes into an untwisted
and a twisted part:
\bdm S_E [G,B,X,Y] \equiv S_E [G,B,X] + S_E [G,B,Y] \; .
\edm
Hence any correlation function factorizes
into a pure orbifold (i.e.\ without
fixed direction) and a toroidal piece:
\be \ba{rcl} \lf\lng O_1 [X,Y] \ldots O_N [X,Y] \ri\rng
     &=& \ds{\int {\cal D}X O_1 [X] \ldots O_N [X]
     e^{-S_E [G,B,X]} } \nnn
     &\times& \ds{\int {\cal D}Y O_1 [Y] \ldots O_N [Y]
     e^{-S_E [G,B,Y]} } \; .
\label{fix} \ea \ee
The calculation of the first factor has been described in the previous
sections for non--excited twist fields. The second factor may generally
be derived from correlators of pure exponential vertex operators
depending only on the coordinate fields $Y$. These
can be obtained via well known methods.

\sect{Duality symmetries}
\label{sec6}

In the preceding part we derived the orbifold couplings with
the full moduli dependence. This gives us the possibility to study their
behaviour under modular transformations which are symmetries of the
spectrum~\cite{dhs,ejn,michal,michal2}. In particular, we investigate
the group generated by
duality symmetries defined in \cite{michal}. This group is not the full
modular symmetry group of orbifold compactifications, but it includes the
well known $R \rightarrow 1/R$ duality transformation.
In this section we define the action of duality on twist fields in such a
way that they balance the transformation of the
twist--anti--twist annihilation
couplings. In the following section we check that this twist field
transformation leads also to invariant Yukawa
couplings and thus completes
the proof of duality invariance of the theory.

Again we will not consider those couplings involving fields
which belong to fixed directions. However, given the discussion in
section~\ref{fixtori} the generalization to these
cases is straightforward.
Equation \req{fix} shows that these couplings possess product structures
concerning twisted and non--twisted directions and solutions
to eq.~\req{wmat} (see below) have the same product form. The latter
becomes explicit when going to the coordinate basis.
This does not mean, however, that the moduli space takes necessarily a
product structure. In other words modular transformations in both factors
might be related.

The group of duality transformations considered in this paper is
generated by transformations of the form
\be
\ba{ccccc}
\ds{\m} &\longrightarrow& \m^\prime &=& W \w \nonumber \\
\ds{\w} &\longrightarrow& \w^\prime &=& W^{\ast} \m \nonumber \\
\ds{g \pm b} &\longrightarrow& g^\prime \pm b^\prime &=&
\ds{W \frac{1}{g \pm b} W^t} \; ,
\ea
\label{duality}
\ee
where hatted quantities refer to the lattice basis, i.e.\ $v=e \w$ and
$p=e^{\ast} \m$. We also have $g=e^t G e$, $b=e^t B e$ and
$Q=e^{-1} \Theta e$, so that $Q$ acts on winding numbers and $Q^\ast
={Q^t}^{-1}$
on momentum quantum numbers. $W$ is a $GL(d,\ZZ)$ matrix. If one
chooses $W$ as a solution of
\be
WQ=Q^\ast W ,
\label{wmat}
\ee
then the transformation \req{duality} is modular~\cite{michal}.

The twist--anti--twist annihilation coupling
\req{ssv} expressed with above
lattice basis quantities reads
($\hat{P}_{R/L} \equiv e^* P_{R/L} =$ $\hat{p}$ $\pm$ $g\hat{v}$
$-$ $b\hat{v}$)
\be
 C_{\f_2,\f_1;\m,\w}^k =\g e^{-2 \pi i\f_2^t \m}\
 e^{i \pi \m^t \frac{1}{1-Q^k} \w}\
 \delta_{\w \in (1-Q^k)(\f_{21}+\ZZZ^d)}\; .
\ee

Under the set of transformations (\ref{duality})
this expression transforms
non--trivially. However, as mentioned above,
an additional unitary redefinition
of the twist fields render the theory invariant~\cite{nilles}:
\be \tilde{\sigma}^{+k}_{\f_{q}}=\frac{1}{\sqrt{N_{k}}}
    \sum_{a=1}^{N_{k}}
    e^{2\pi i\f_{a}^{\ast^{t}} (1-Q^k)\f_{q}}\
    \sigma^{+k}_{W^{\ast}\f_{a}^{\ast}}\ \ \ ,\ \ \
    \tilde{V}^{\rm inv}_{\m,\w}=V^{\rm inv}_{\m',\w'}
    e^{i \pi \m^{t}\w}\ .  \label{sr}
\ee
Here $N_k$ denotes the number of fixed points $\f_{a}^{\ast}$ under
$Q^{\ast k}$.
The additional phase in the transformation of exponential type vertex
operators in \req{sr} does not affect their correlation functions as it
can be shown to be equivalent to
a redefinition of cocycle operators. Since the twisted sector string
couplings $g_k(P_L,P_R)$ only depend on
the conformal dimensions of untwisted vertex
operators, they are guaranteed to be invariant, because of the
duality symmetry of the spectrum.

Using the rules (\ref{duality})
and (\ref{sr}) a lengthy but straightforward
calculation (for details see \cite{stieber}) shows that
\bdm
\ds{\tilde{C}^k_{\f_2,\f_1;\m^\prime ,\w^\prime}(g',b')} =
C^k_{\f_2,\f_1;\m,\w}(g,b)\ ,
\edm
where $\tilde{C}^k$ denotes the coupling defined through the transformed
fields in \req{sr}.
For this we used the
identity $\frac{1}{1-Q^k}+\frac{1}{1-Q^{-k}}=1$,
the fact that  $(1-Q^k)\f \in \ZZ^d$ and the projector property \cite{rp}
\be
\frac{1}{N_{k}}\sum_{b=1}^{N_{k}} e^{-2\pi i \f_{b}^{\ast^{t}} \w}
= \left\{ \begin{array}{ll} 1 & \mbox{if $\w \in (1-Q^k)\ZZ^d$} \\ 0 &
\mbox{otherwise,}
\end{array}
\right.
\label{jp}
\ee
where $\f_b^\ast$ is fixed under ${Q^\ast}^k$.

Since from (\ref{wmat}) it follows that $W^{\ast}{Q^\ast}^k=Q^k\
W^{\ast}$,
we see that any fixed point $\f^{\ast}$ under ${Q^\ast}^k$
satisfies the equation
\be
\ds{Q^k \ (W^{\ast} \f^{\ast})= (W^{\ast}\ \f^{\ast}) +\ZZ^d},
\label{Wf}
\ee
which means that $W^{\ast} \f^{\ast}$ is a fixed point under $Q^k$.
Likewise, $W f$ is fixed under ${Q^{\ast}}^k$ whenever $\f$ is fixed under
$Q^k$. In fact, replacing $W^{\ast} \f^{\ast}$
in (\ref{sr}) by $\f$ gives
\be
\ds{\tilde{\sigma}^{k}_{\f_{q}}=
\frac{1}{\sqrt{N_{k}}}\sum_{a=1}^{N_{k}}
e^{2\pi i \f_{a}^t (1-Q^{\ast^k})W \f_{q}} \  \sigma^{k}_{\f_{a}}},
\label{sra}
\ee
showing that twist fields defined w.r.t.\ dual lattice fixed points, are
just linear combinations of the ones related to $\Lambda$ itself. This is
the essential observation of the duality
proof presented here. Note also the
important role of the matrix $W$, which is necessary to pass from fixed
points to dual ones.

\sect{Yukawa couplings as automorphic functions}
\label{sec7}

We will now discuss the transformation
properties of the Yukawa couplings.
Since we fixed the unitary redefinition of twist fields in the
previous section\footnote{Note however, that we are still free to
perform a unitary phase multiplication leaving the annihilation couplings
untouched.}, invariance of the Yukawa couplings
must be due to a mathematical
identity~\cite{nilles}. This turns out to be a Poisson resummation
exchanging basically winding and momentum lattices.

We start by recasting the exponent in equation (\ref{sss}) to read
\be
v^t(G+B)Lv = v^t(GZ+iB)Av,
\ee
where the matrices $A$ and $Z$ are
block diagonal, with the non--vanishing elements
\bea
&\ds{A_{2j-1,2j}=-A_{2j,2j-1}=\frac{1}{\lambda_j}} ,& \\
&\ds{Z_{2j,2j-1}=-Z_{2j-1,2j}=1} .&
\eea
In terms of lattice basis quantities the three twist field correlator
\req{sss} reads:
\be
Y^{k,l}_{\f_a,\f_b,\f_c}[g,b]= (2\pi)^{\fc{d}{4}} (\det g)^{\frac{1}{4}}\
\lf[\prod_{j=1}^{d/2} \Gamma_{k_j,l_j} \ri]^{\hal}
\sum \limits_{\w \in \fc{1-Q^{k+l}}{1-Q^k}
(\f_{cb} +\tau_b +\fc{1-Q^k}{1-Q^{\phi}}\ZZZ^d)}
e^{-2\pi \w^t(gz+ib)a \w}\ ,
\label{yukawa}
\ee
where $z:=e^{-1}Ze$ and
$a:=e^{-1}Ae=-\hal \fc{(1-Q^k)(1-Q^l)}{1-Q^{k+l}}$.

For simplicity we will restrict ourselves to cases,
where in every complex
subspace the condition $ 0 < k_i + l_i < 1 $ holds.  Inserting the
rules (\ref{duality}) and (\ref{sr})
into expression (\ref{yukawa}) yields
\bea
\ds{\tilde{Y}^{k,l}_{\f_a,\f_b,\f_c}[g',b']} &=&
\ds{(2\pi)^{\fc{d}{4}}\; ({\det} g')^{\frac{1}{4}}\ \
\lf[\prod_{j=1}^{d/2}\Gamma_{k_jl_j}\ri]^{\hal}\
\frac{1}{\sqrt{N_{k}N_{l}N_{k+l}}}} \nnn
&& \hspace{-2.5cm}\ds{\times\;\;\sum_{p,q,r=1}^{N_k,N_l,N_{k+l}}
e^{2\pi i [ \f_{p}^{\ast^{t}} (1-Q^k)\f_{a} +
\f_{q}^{\ast^{t}} (1-Q^l)\f_{b} +
\f^{\ast^{t}}_{r}(1-Q^{-(k+l)})\f_{c}] }
\sum_{W^{\ast} \m \in \cal{R}}
e^{-2\pi i  \m^t \frac{1}{g+b}(gz-ib) \fc{1}{g-b} a\m}}\; ,
\label{Y28}
\eea
with ${\cal R}=\fc{1-Q^{k+l}}{1-Q^k}
\lf[ W^{\ast} (\f_{rq}^{\ast} + \tau_{q})
+\fc{1-Q^k}{1-Q^{\phi}}\ZZ^d\ri]$.
The couplings appearing on the rhs obey the selection rule
(\ref{sg}), which we represent in the form
\be \label{selrule}
(1-Q^k)W^{\ast}\f_p^{\ast}+Q^k(1-Q^l)W^{\ast}\f_q^{\ast}-(1-Q^{k+l})
W^{\ast}\f_r^{\ast}=(1-Q^\phi) W^\ast \tau .
\ee
This can be taken into account by multiplying each summand on the rhs of
(\ref{Y28}) (labeled by $(p,q,r)$) with
\bea
&&\hspace{-3cm}\ds{\fc{1}{N_{\phi}}\sum_{j=1}^{N_{\phi}}
e^{2\pi i \f_j^t
[(1-Q^{\ast^k})\f^{\ast}_p+Q^{\ast^k}
(1-Q^{\ast^l})\f^{\ast}_q-(1-Q^{\ast^{k+l}})\f^{\ast}_r]}} \nnn
\hspace{2cm}&=&
\ds{\lf\{\ba{clc} 1 & {\rm if\ }
{\rm the\ dual\ fixed\ points\ satisfy\ \req{selrule},} \\
0 &  {\rm otherwise.}
\ea \ri.}
\eea

Before continuing let us remark that the matrices $a$ and $z$ both
commute with the background, which is most easily seen in the coordinate
basis. Using this, as well as $z^2 = -1$, one can prove the identity
\be
   \ds{\m^t \frac{1}{g+b}(gz-ib)\frac{1}{g-b} \m
   = -\m^t \frac{1}{(gz+ib)} \m}\; .
\ee
One also needs the equality
\be \sqrt{\det a^{-1}}=
    2^{d/2}\ \sqrt{\frac{N_{k+l}}{N_k\  N_l}}\; .
    \label{euler}
\ee

We are now prepared to perform the Poisson resummation, i.e.\ we use
the well known formula
\be
\sum_{\m \in \ZZZ^d} e^{-\pi(\m+\hat{\varphi})^t a (\m+\hat{\varphi})}
=\fc{1}{\sqrt{\det\ a}} \sum_{\w\in \ZZZ^d}e^{-\pi \w^t a^{-1} \w +2\pi i
\w^t\hat{\varphi}}\ .
\ee
This yields the intermediate result
\bdm
\ba{lrl}
\ds{\tilde{Y}^{k,l}_{\f_a,\f_b,\f_c}[g',b']} &=&
\ds{(2\pi)^{\fc{d}{4}} ({\rm det} g)^{\frac{1}{4}}\lf[\frac{{\det}(gz+ib)}
{{\rm det}(gz-ib)}\ri]^{\fc{1}{4}}\ \lf[\prod_{i=1}^{d/2}
\Gamma_{k_il_i}\ri]^{\hal}\sum_{\w \in \ZZZ^d}
e^{\fc{\pi}{2} \w^t {a'}^{-1}(gz+ib) \w}} \nnn
&&\hspace{-2cm}\times\;\;\; \ds{\sum_{j=1}^{N_\phi}
\fc{1}{N_k}\sum_{p=1}^{N_k}e^{2\pi i \f_{p}^{\ast^{t}}
[(1-Q^k)\f_{a}-(1-Q^k)\f_j-\fc{1-Q^{bl}}{Q^{-k}(1-Q^{-l})}\w]}}\nnn
&&\hspace{-2cm}\times\;\;\; \ds{\fc{1}{N_l}\sum_{q=1}^{N_l}
e^{2\pi i \f_{q}^{\ast^{t}}
[(1-Q^l)\f_{b}-(1-Q^l)Q^{-k}\f_j+\fc{1-Q^{-\phi}}
{1-Q^{-k}}\w-\fc{1-Q^{bl}}{1-Q^{-k}}\w]}}\nnn
&&\hspace{-2cm}\times\;\;\;\ds{\fc{1}{N_{k+l}}\sum_{r=1}^{N_{k+l}}
e^{2\pi i\f^{\ast^t}_{r}
[-(1-Q^{k+l})\f_{c}+(1-Q^{k+l})\f_j-\fc{1-Q^{-\phi}}
{1-Q^{-k}}\w+\fc{1-Q^{bl}}{(1-Q^{-k})Q^{-k}}
\fc{1-Q^{-(k+l)}}{1-Q^{-l}}\w]}}\; ,
\ea
\edm
where $a'=a \det \lf|\fc{1-Q^{k+l}}{1-Q^{\phi}}\ri|^2$.
The sums over $p$, $q$ and $r$ are just
projectors restricting the allowed winding vectors and fixing $f_j$.
This shows that only one value of $j$ gives a non vanishing contribution,
so that the sum over $j$ can also be performed.
After rewriting the exponential in its original form we arrive at the
result
\bea
\ds{\tilde{Y}^{k,l}_{\f_a,\f_b,\f_c}[g',b']} &=& \ds{(2\pi)^{\fc{d}{4}}
(\det g)^{\frac{1}{4}}\lf[\frac{\det (gz+ib)}
{\det (gz-ib)}\ri]^{\frac{1}{4}}\
\lf[\prod_{j=1}^{d/2}\Gamma_{k_jl_j}\ri]^{\hal}}\nnn
&\times&\ds{\sum_{\w \in \ZZZ^d} e^{-2\pi \w^t (gz+ib)a \w}\
\delta_{\w \in \fc{1-Q^{k+l}}{1-Q^l}(\f_c-\f_a+\ZZZ^d)}
\delta_{\w \in \f_b-\f_a+\ZZZ^d}}\nnn
&=& \ds{\lf[\frac{\det (gz+ib)}{\det(gz-ib)}\ri]^{\frac{1}{4}}
Y^{k,l}_{\f_a,\f_b,\f_c}[g,b]}\; .
\eea
We see that we encounter a phase factor due
to the antisymmetric background
field, which has to be canceled by a unitary phase transformation of the
twist fields.
The fact that this phase can be written in a
covariant form is a consequence
of our assumption concerning the twist
eigenvalues $ 0 < k_j + l_j < 1 $. In the general case
with arbitrary $k_j$ and $l_j$, however, the corresponding factor is an
explicit product over the complex dimensions. Thus, for the theory
to be invariant, the twist fields have to be transformed according to
\be
{\sigma}^{\pm k}_{\f_{q}} \mapsto \tilde{\sigma}^{\pm k}_{\f_{q}}=
\frac{1}{\sqrt{N_{k}}}\prod_{j=1}^{d/2}
\lf[\frac{{\det_j} (GZ-iB)} {{\det_j}(GZ+iB)}\ri]^{\pm \frac{1}{4}
(1-2k_j)}
\sum_{a=1}^{N_k} e^{2\pi i\f_{a}^{t} (1-{Q^{\ast}}^{\pm k}) W \f_{q}}\;
\si^{\pm k}_{\f_a}\; ,
\label{correct}
\ee
where the index $j$ refers to the respective complex subspace.
Phrased in mathematical terms the invariance of the Yukawa couplings
translates into the statement that it represents an automorphic function
of the relevant modular symmetry group.

The rather lengthy expressions for the couplings and
duality transformations simplify considerably in cases, where the twist
order $N$ is prime. This can be traced back to the fact that in this case
any power $1< k < N$ of the twist matrix
could be chosen equivalently to be the
{\em defining\/} one. From this observation it follows immediately that
all twist sectors of the theory are equivalent.
The Yukawa couplings (\ref{sss}) in this case can be written in the form
\be
Y_{f_a,f_b,f_c}^{k,l}[G,B]=(2\pi)^{\fc{d}{4}}\lf[V_{\Lambda}
\prod_{j=1}^{d/2} \Gamma_{k_j,l_j}\ri]^{\hal}
\sum_{v \in f_b-f_a+\Lambda}  e^{-2\pi v^t(G+B)Lv}\ ,
\label{psss}
\ee
showing that it only depends on the differences of pairs of
fixed points. In fact, the selection rule
(\ref{sg}) is restrictive enough
to enable us to denote the Yukawa couplings solely by one quantity
referring to the fixed point structure (in addition to the twist sector
configuration).
With the definition $Y_{f_a,f_b,f_c}^{k,l} \equiv Y_{b-a}^{k,l}$
we can write the Yukawa couplings w.r.t.\ the new
background in terms of the ones measured at the original values as
\be  Y_{d}^{k,l}[g',b']=\frac{1}{\sqrt{N}}
     \lf[\frac{\det (gz-ib)}{\det (gz+ib)}\ri]^{\frac{1}{4}}
     \sum_{j=1}^{N} e^{-2 \pi i \f_j^{\ast^t}
     \frac{(1-Q^{\ast^k})(1-Q^{\ast^l})}{1-Q^{\ast^{k+l}}}W\f_d}\
     Y^{k,l}_{j}[g,b]\ .
\ee
Thus, comparing with (\ref{correct}) we see that they transform in a way
very similar to the twist fields themselves,
showing explicitly how Yukawa couplings form a representation of the
group of modular  transformations\footnote{Recall that the physical
Yukawa couplings are in general defined through twist invariant
combinations of twist fields.
Compare the remark following eq.~\req{twistinv}.}.

\vspace{1 cm}
\ \\
\bf{Acknowledgements}: \normalsize It is a
pleasure for us to thank Albrecht
Klemm and Hans--Peter Nilles for many useful
discussions and comments. We are especially grateful to J\"urgen
Lauer for initiating this work as well as
for some important contributions.

\vspace{ 1 cm}

\newpage

\section*{Appendix}

\appendix

\sect{The restricted instanton summation range}
\label{sets}

In this appendix we derive the relation (\ref{sumrange})
\be F_1 \cap\lf( -F_3\ri) = (1-\Th^k)(f_{21}+\tilde{\La})
\ee
with
\be \tilde{\La} := -T_1 + \fc{1-\Th^l}{1-\Th^\phi} \La \; ,
\ee
for the restricted instanton summation range of the four twist field
correlator as a solution of the consistency condition $v_3$ $=$ $-v_1$.
$\phi$ denotes the greatest common divisor of $k$ and $l$.

With the help of the space group selection
rule (\ref{sgsr1}) an arbitrary
$v_1 = (1-\Th^k)(f_{21}+\alpha ) \in  F_1\; ;\;\; \alpha\in\La$
can be written as
\bdm v_1 = - (1-\Th^l) \lf[ f_{43} - T_2
     - \fc{1-\Th^k}{1-\Th^l}(\alpha + T_1) \ri] \; .
\edm
For this object to be an element of the set $-F_3$ we have to restrict
$\alpha$ $\in$ $\La$ in such a way, that
\be \fc{1-\Th^k}{1-\Th^l}(\alpha + T_1) \in \La \; .
    \label{condalpha}
\ee
We want to show, that the set of solutions is given by $\tilde{\La}$.
With the help of the identity
$\fc{1-\Th^k}{1-\Th^\phi} = \sum_{j=0}^{k_0 - 1}\Th^{j\phi}$,
where $k_0$ $\in$ $\ZZ$ is defined via $k$ $=$ $k_0\phi$,
it is obvious, that any $\alpha$ $\in$ $\tilde{\La}$ fulfills
(\ref{condalpha}). In order to show that $\tilde{\La}$ actually comprises
{\em all}
solutions of (\ref{condalpha}) we first write
$\alpha$ $=$ $-T_1$ $+$ $\la$
for some $\la$ $\in$ $\La$. Next we decompose the winding lattice
into fixed points cosets:
$\La = \bigcup_{f^l \in {\cal F}_l} (1-\Th^l)(f^l + \La )$,
where ${\cal F}_l$ denotes the set of all $f^l$,
for which $\tilde{f}^l$ $=$
$2\pi f$ is fixed under
$\Th^l$, i.e.\ $(1-\Th^l)\tilde{f}^l$ $\in$ $2\pi\La$. Hence we may write
$\alpha$ $=$ $-T_1$ $+$ $(1-\Th^l)(f^l + \rho )$ for some $f^l$ $\in$
${\cal F}_l$ and $\rho$ $\in$ $\La$. Since ${\cal F}_\phi$
$\subset$ ${\cal F}_l$ but generally ${\cal F}_\phi$ $\neq$ ${\cal F}_l$,
we can distinguish two cases for $f^l$ $\in$ ${\cal F}_l$:
\begin{enumerate}
 \item $f^l \in {\cal F}_\phi$: In this case $\alpha$ trivially fulfills
       (\ref{condalpha}), since each fixed point under $\Th^\phi$ is
       also fixed under $\Th^k$.
 \item $f^l \nin {\cal F}_\phi$:
       Now $\alpha$ does not obey (\ref{condalpha}),
       since $f^l$ cannot be fixed under $\Th^k$:
       Any $f^l$ fixed under $\Th^l$ {\em and} $\Th^k$ is also fixed
       under $\Th^{nk+ml}$ for arbitrary $n,m$ $\in$ $\ZZ$.
       Since the greatest common divisor of two integers $k$ and $l$
       may always be written in the form $\phi$ $=$ $nk$ $+$ $ml$,
       for certain $n,m$ $\in$ $\ZZ$, we arrive at the conclusion, that
       $f^l$ then must be fixed under $\Th^\phi$ as well, which
       contradicts the assumption $f^l \nin {\cal F}_\phi$.
\end{enumerate}
Hence we have proven, that
\bdm \alpha \in -T_1 + \fc{1-\Th^l}{1-\Th^\phi}
     \bigcup_{f^\phi \in {\cal F}_\phi}
     (1-\Th^\phi )(f^\phi + \La ) =
     \tilde{\La} \; .
\edm

Analogously one easily shows, that for
$v_3 = (1-\Th^l)(f_{43}+\beta )$ $\in$ $F_3$; $\beta$ $\in$ $\La$
to be an element of $-F_1$, one has to require that
$\beta \in \Lambda':=-T_2 + \fc{1-\Th^k}{1-\Th^\phi} \La $.
This leads to the alternative representation
$F_1 \cap\lf( -F_3\ri) = -(1-\Th^l)(f_{43}+\La^\prime )$.

Note that the sets $\tilde{\La}$ and $\La^\prime$
are unambiguously determined,
in the sense that they do not depend on the choice of $T_1$, $T_2$,
as long as the latter solve (\ref{sgsr1}).
Indeed, a different choice of $T_1$, $T_2$ (while keeping
$f_{21}$, $f_{43}$ fixed) can be readily absorbed by suitable shifts
of the $\Lambda$ factors present in the definitions of
$\tilde{\La}$ and $\La^\prime$.
Another ambiguity is met if the fixed points $f_{21}$,
$f_{43}$ are displaced by lattice vectors. Clearly, with the same
shift applied to $T_1$, $T_2$ in (\ref{sgsr1}) compensation sets in.

\sect{Direct calculation of Yukawa couplings}
\label{direct}

In this section we sketch a direct calculation of the world sheet
instanton contribution to the Yukawa coupling.
This method does not rely on
factorizing the four point correlation function and therefore
the correct normalization of the Yukawa couplings cannot be deduced.
Nevertheless it serves as an interesting
cross--check for our results obtained in section \ref{YYY}.

One begins by examining the global monodromy conditions.
As in the case of the four point function, we have to consider closed
net--twist zero loops. A possible minimal choice of loops,
which determine the full global monodromy, is given by $C_1$, $C_2$ and
$C_3$, which are defined to encircle the pairs of twist fields
$(\si_{f_2}^{l}(1,1),\si_{f_3}^{-(k+l)}(0,0))$,
$(\si_{f_2}^{l}(1,1),\si_{f_1}^{k}(z_\infty ,\zb_\infty ))$ and
$(\si_{f_3}^{-(k+l)}(0,0),\si_{f_1}^{k}(z_\infty ,\zb_\infty ))$,
respectively.
The corresponding global monodromy conditions are given by (\ref{clmond}) with
monodromy vectors
\bea
 v_1 &\in& (1-\Th^{rl})(f_{23}+ \al_1) \nnn
 v_2 &\in& (1-\Th^{pk})(f_{21}+\al_2)  \nnn
 v_3 &\in& (1-\Th^{mk})(f_{31}+\al_3)
\eea
and $ \ds{\al_1, \al_2, \al_3 \in \Lambda} $. The integer numbers
$p,q,m,n,r$ and $s$ are defined to satisfy
\bea
 \ds{pk} &=& \ds{ql}  \nnn
 \ds{rl} &=& \ds{s(k+l)} \nnn
 \ds{mk} &=& \ds{n(k+l)} \; .
\eea
They give the number of times the corresponding points are encircled
to ensure that the loops close. For the case $k=l$ only one loop is
required, but in the general case studied
here it turns out to be necessary to consider {\em two} loops.

The classical solutions are
\be
\ba{cll}
\ds{\p Y^i_{cl}(z)} &=& \ds{a^i \Omega_{k_i,l_i}(z):=a^i
 (z-z_{\infty})^{-(1-k_i)}z^{-(1-l_i)}
(z-1)^{-(k_i+l_i)}} \\[4mm]
\ds{\bar \p Y^i_{cl}(\bar z)}
&=& \ds{b^i \Omega_{1-k_i,1-l_i}(\bar z):= b^i
 (\bar z-\bar z_{\infty})^{-k_i} \bar z^{-l_i}
(\bar z-1)^{-(1-k_i-l_i)}}
\ea
\ee
As in the case of the four--point correlation function,
the global monodromy
conditions \req{clmond} result in an over--determined system
of equations for the coefficients  $a^i, b^i$.
These equations can be found by
calculating the contour integrals \cite{klein}
\be
\lim_{z_\infty \ra \infty} (-z_\infty)^{1-k_i}\ a^i \oint_{C_j} dz \
\Omega_{k_i,l_i}(z)=
2\pi v_j^i \ \ ,\ \ j=1,2,3\ .
\label{tmc}
\ee
They do not only determine the coefficients $a^i, b^i$, but also yield
restrictions on the summation range. These can be written
in the form
\bea
\ds{f_{31}+\al_3} &=& \ds{\fc{\Th^k(1-\Th^l)}{1-\Th^k}(f_{23}+\al_1) }
\nnn \\
\ds{f_{21}+\al_2} &=& \ds{\fc{1-\Th^{k+l}}{1-\Th^k}(f_{23}+\al_1)\; .}
\label{a1a2a3}
\eea
Thus, the $\alpha_i$ do not run over the
whole lattice, but over a subset.
One can check, that the equations \req{a1a2a3} are solved by
\bea \ds{\alpha_1} &\in& \ds{-\tau_2 +
     \frac{1-\Theta^k}{1-\Theta^\phi} \Lambda } \nnn
     \ds{\alpha_2} &\in& \ds{\tau_1 -\tau_2 +
     \frac{1-\Theta^{k+l}}{1-\Theta^\phi} \Lambda } \nnn
     \ds{\alpha_3} &\in& \ds{\tau_1 +
     \frac{\Th^k(1-\Th^l)}{1-\Theta^\phi} \Lambda \; .}
\eea
Here $\tau_1, \tau_2$ denote lattice vectors appearing in the
selection rule for the three point functions:
\eq
(1-\Th^k)(f_1-\tau_1)+\Th^k (1-\Th^l) (f_2-\tau_2) = (1-\Th^{k+l})f_3\; .
\qe
One may choose any of the above summation variables to represent the
instanton sum. The representation  used
in section \ref{YYY} corresponds to  the choice  $\alpha_2$.
The corresponding contribution in the classical action
(\ref{action1})  are determined  to be
\be
\ba{clcr}
&\ds{\frac{1}{16 \sin^2(\pi rl_i)}\ \lf|\frac{\sin(\pi l_i)
\sin[\pi(k_i+l_i)]}{\sin(\pi k_i)}\ri|
|v_1^i|^2}& \\[4mm]
&\ds{\frac{1}{16 \sin^2(\pi pk_i)}\
\lf|\frac{\sin(\pi k_i) \sin(\pi l_i)}
{\sin[\pi(k_i+l_i)]}\ri|
|v_2^i|^2}&\\[4mm]
&\ds{\frac{1}{16 \sin^2(\pi mk_i)}\ \lf|\frac{\sin(\pi k_i)
\sin[\pi(k_i+l_i)]}{\sin(\pi l_i)}\ri|
|v_3^i|^2\ ,}&
\ea
\ee
respectively.

\end{document}